\documentclass{article}

\usepackage{authblk}
\usepackage[square]{natbib}
\usepackage{yhmath}
\usepackage{mathtools, amsmath, amsthm, amssymb, amsbsy, amsfonts, amscd}
\usepackage{euscript}
\usepackage{bm}
\usepackage{breqn}
\usepackage{xcolor, soul}
\usepackage{empheq}
\usepackage{rotating}
\usepackage{enumerate}
\usepackage{enumitem}
\usepackage{bbold}

\usepackage{ftnxtra}
\usepackage{fnpos}
\usepackage{appendix}

\usepackage{graphicx}
\usepackage{pstool}
\usepackage{psfrag}

\usepackage{comment}

\numberwithin{equation}{section}

\theoremstyle{plain}	
 \newtheorem{thm}{Theorem}[section]

 \newtheorem{prop}[thm]{Proposition}
\theoremstyle{definition}	
 \newtheorem{defi}{Definition}[section]
 \newtheorem{remark}{Remark}[section]

\usepackage{caption2}

\usepackage{hyperref}
\hypersetup{colorlinks=true, linkcolor=blue}
\hypersetup{colorlinks=true,citecolor=blue}

\DeclareMathAlphabet{\mathpzc}{OT1}{pzc}{m}{it}

\usepackage{mathrsfs}

\definecolor{lighter_purple_mathematica}{rgb}{0.6666666666,0.33333333333,0.666666666666}

\makeatletter
\newsavebox{\@brx}
\newcommand{\llangle}[1][]{\savebox{\@brx}{\(\m@th{#1\langle}\)}%
  \mathopen{\copy\@brx\mkern2mu\kern-0.9\wd\@brx\usebox{\@brx}}}
\newcommand{\rrangle}[1][]{\savebox{\@brx}{\(\m@th{#1\rangle}\)}%
  \mathclose{\copy\@brx\mkern2mu\kern-0.9\wd\@brx\usebox{\@brx}}}%
\let\oldabs\abs
\def\abs{\@ifstar{\oldabs}{\oldabs*}}
\makeatother

\usepackage{accents}

\newcommand{\g}{\mathbf{g}}
\newcommand{\G}{\mathbf{G}}
\newcommand{\Go}{\mathring{\G}}

\newcommand{\F}{\mathbf{F}}
\newcommand{\Fe}{\accentset{e}{\F}}
\newcommand{\Fa}{\accentset{a}{\F}}
\newcommand{\Fn}{\accentset{n}{\F}}
\newcommand{\Fth}{\accentset{\scalebox{0.4}{\(\Theta\)}}{\F}}

\newcommand{\cF}{\mathrm F}

\newcommand{\sharpo}{\mathring\sharp}
\newcommand{\flato}{\mathring\flat}

\newcommand{\Ds}{\accentset{s}{\mathbf{D}}}
\newcommand{\Da}{\accentset{a}{\mathbf{D}}}

\newcommand{\pe}{\accentset{e}{p}}
\newcommand{\pn}{\accentset{n}{p}}
\newcommand{\Je}{\accentset{e}{J}}
\newcommand{\Jn}{\accentset{n}{J}}

\newcommand{\Bn}{\accentset{n}{\mathbf{B}}}
\newcommand{\be}{\accentset{e}{\mathbf{b}}}
\newcommand{\Bth}{\accentset{\scalebox{0.4}{\(\Theta\)}}{\mathrm B}}

\usepackage{dsfont}

\usepackage{empheq}

\date{20 August 2025}

\begin{document}

\title{\textbf{A Generalized Coleman-Noll Procedure and the Balance Laws of Hyper-Anelasticity}}

\author[1]{Souhayl Sadik\thanks{Corresponding author e-mail: sosa@mpe.au.dk}}
\author[2,3]{Arash Yavari}
\affil[1]{\small \textit{Department of Mechanical and Production Engineering, Aarhus University, 8000~Aarhus~C, Denmark}}
\affil[2]{\small \textit{School of Civil and Environmental Engineering, Georgia Institute of Technology, Atlanta, GA 30332, USA}}
\affil[3]{\small \textit{The George W. Woodruff School of Mechanical Engineering, Georgia Institute of Technology, Atlanta, GA 30332, USA}}

\maketitle


\begin{abstract}
It is known that the balance laws of hyperelasticity (Green elasticity), i.e. conservation of mass and balance of linear and angular momenta, can be derived using the first law of thermodynamics and by postulating its invariance under superposed rigid body motions of the Euclidean ambient space---the Green-Naghdi-Rivlin theorem. In the case of a non-Euclidean ambient space, covariance of the energy balance---its invariance under arbitrary time-dependent diffeomorphisms of the ambient space---gives all the balance laws and the Doyle-Ericksen formula---the Marsden-Hughes theorem. It is also known that, by assuming the balance laws, and positing the first and second laws of thermodynamics, the Doyle-Ericksen formula can be derived\textemdash the Coleman-Noll procedure.
Traditionally, the first law of thermodynamics combined with an invariance assumption has been used to derive the balance laws, while the second law has served to constrain constitutive equations. In this paper, we explore how the balance laws themselves can be derived directly from thermodynamic principles.
We accomplish this via a generalization of the Coleman-Noll procedure: it is shown that the Doyle-Ericksen formula as well as the balance laws for both hyperelasticity and hyper-anelasticity can be derived using the first and second laws of thermodynamics without assuming any (observer) invariance.
\end{abstract}

\begin{description}
\item[Keywords:] Coleman-Noll procedure, nonlinear elasticity, hyperelasticity, anelasticity, hyper-anelasticity, balance laws, laws of thermodynamics.
\item[Mathematics Subject Classification] 74A15 $\cdot$ 74B20
\end{description}

\tableofcontents

\section{Introduction}
\label{S:intro}

There are several approaches to derive the balance laws in field theories, particularly in elasticity.
One approach is the Lagrangian field theory (variational approach) of elasticity. In this method, Hamilton’s principle is written for an elastic body, leading to the balance of linear momentum as the Euler-Lagrange equations of the variational principle. The balance of angular momentum is then derived using Noether’s theorem, which relates the invariance of the Lagrangian density under ambient space rotations to angular momentum conservation. The balance of energy corresponds to the invariance of the Lagrangian density under time shifts.
The second approach is the Hamiltonian mechanics formulation. In this framework, nonlinear elasticity is represented in phase space, where Hamilton’s equations give the balance laws \citep{Simo1988}.
More specifically, the governing equations of nonlinear elasticity can be derived using the canonical Hamilton equations $\dot{f}=\{f,H\}\,$, where $\{\}$ is the Poisson bracket of nonlinear elasticity, $H$ is the Hamiltonian of nonlinear elasticity and $f$ is an arbitrary scalar function.

Another approach to derive the balance laws of nonlinear hyperelasticity is to start from an energy balance (the first law of thermodynamics) and postulate its invariance under superposed rigid body motions of the ambient space (observer invariance). This idea is due to \citet{Green64} in the context of Euclidean ambient spaces (Green-Naghdi-Rivlin theorem). More specifically, \citet{Green64} postulated the balance of energy and its invariance under superposed translational and rotational motions of the Euclidean ambient space. 
A different version of this theorem is due to \citet{Noll1963} who thought of the superposed motions passively as time-dependent coordinate charts for the Euclidean ambient space.\footnote{Recall that a matrix can be regarded as either a linear transformation (active) or representing a change of basis (passive).} Effectively, \citet{Green64} viewed superimposed motions actively, whereas \citet{Noll1963} viewed them passively.
The invariance idea was subsequently extended to hyperelasticity (Green elasticity) with Riemannian ambient space manifolds by \citet{HuMa1977}; they postulated the invariance of the balance of energy under arbitrary diffeomorphisms of the ambient space---\emph{covariance} of the energy balance. \citet{HuMa1977} showed that covariance of the energy balance gives all the balance laws of hyperelasticity and the Doyle-Ericksen formula \citep{doyle1956nonlineari}\footnote{It is worth emphasizing that both in \citep{Green64} and \citep{HuMa1977}, it was assumed that the body is made of a material that has an underlying energy function, i.e. they restricted themselves to hyperelasticity. 
Similar invariance arguments can be used to derive the balance laws of anelasticity, provided that there exists an underlying energy function, e.g., \citep{Yavari2010}.} (see also \citep{MarsdenHughes1983,YaMaOr2006}).
It is worth noting that the relationship between the covariance of the energy balance and the Lagrangian field theory of elasticity was explored in detail in \citep{YavariMarsden2012}.\footnote{A widely used approach for deriving the balance laws, particularly in computational mechanics, is the principle of virtual work (or virtual power) \citep{Antman1979,Maugin1980}. It can be shown that the principle of virtual work can be derived from the balance of energy and its covariance \citep{MarsdenHughes1983}.}

The second law of thermodynamics imposes constraints on the constitutive equations. In the classical \citet{coleman1963} procedure, it is assumed that the balances of linear and angular momenta are already satisfied. Subsequently, the second law of thermodynamics, in the form of the Clausius-Duhem inequality, places restrictions on the form of the constitutive equations.

Rather than relying on the first law of thermodynamics and the assumption of invariance (or covariance), this paper introduces an extension of the Coleman–Noll procedure \citep{coleman1963}.
We show that one can derive not only the Doyle–Ericksen formula but also the full set of balance laws of hyperelasticity using the first and second laws of thermodynamics alone.
This is accomplished by generalizing the notion of a thermodynamic process—originally introduced by \citet{coleman1963}—to what we call an \emph{extended thermodynamic process}. We then require the Clausius–Duhem inequality to hold for a class of such processes refered to as \emph{admissible}\textemdash processes for which the constitutive assumptions are satisfied at all times throughout the body. The proposed framework naturally extends to models of hyper-anelasticity, highlighting its broader applicability beyond purely hyperelastic setting.

This paper is organised as follows:
In \S\ref{Sec:Kinematics}, we introduce some notational elements and discuss the kinematics framework, in particular we discuss the material geometry for finite elasticity and anelasticity including the Bilby-Kröner-Lee decomposition in the anelastic case. 
In \S\ref{Sec:ThermoElast}, we discuss the thermodynamics of hyperelastic solids and, without making any invariance assumptions, demonstrate that one can derive not only the hyperelastic constitutive equations (as in the classical Coleman-Noll procedure) but also the balance laws of hyperelasticity. This leads to what we refer to as the generalized Coleman-Noll procedure.
In \S\ref{Sec:ThermoAnElast}, we extend our generalized Coleman-Noll procedure to hyper-anelasticity. We further discuss the heat equation and the kinetic equations governing the evolution of anelastic distortions.
We conclude the paper with some final remarks  in \S\ref{Sec:Con}.

\section{Kinematics}
\label{Sec:Kinematics}

\subsection{Kinematic and Mathematical Preliminaries}

Consider a solid body $\mathsf{B}$ represented by an embedded $3$-submanifold $\mathcal{B}$ within the ambient space $\mathcal{S}$.\footnote{For most applications, the ambient space is the three-dimensional Euclidean space, i.e. $\mathcal{S}=\mathbb R^3\,$. However, in general, the ambient space may be curved, e.g., in modeling the dynamics of fluid membranes \citep{Arroyo2009}. See \citep{Yavari2016} for a general framework on elasticity in evolving ambient spaces.} Motion of the body $\mathsf{B}$ is represented by a time-parametrized family of maps $\varphi_t:\mathcal{B}\to\mathcal{C}_t\subset\mathcal{S}\,$, mapping the reference (material) configuration $\mathcal{B}$ of the body to its current (spatial) configuration $\mathcal{C}_t=\varphi_t(\mathcal{B})\,$.
We adopt the following standard convention: objects and indices are denoted by uppercase characters in the material manifold $\mathcal{B}$ (e.g., $X\in\mathcal{B}$), and by lowercase characters in the spatial manifold $\varphi_t(\mathcal{B})$ (e.g., $x=\varphi_t(X)\in\varphi_t(\mathcal{B})$). We consider local coordinate charts on $\mathcal{B}$ and $\mathcal{S}$ and denote them by $\{X^A\}$ and $\{x^a\}\,$, respectively. The corresponding local coordinate bases are denoted by $\{\partial_A=\frac{\partial}{\partial X^A}\}$ and $\{\partial_a=\frac{\partial}{\partial x^a}\}\,$, and their respective dual bases are $\{dX^A\}$ and $\{dx^a\}\,$. We adopt Einstein's repeated index summation convention, e.g.,~$u^i v_i\coloneq \sum_i u^i v_i\,$.

In the ambient space $\mathcal S\,$, given a vector $\mathbf{u}\in T_x\mathcal{S}$ and a 1-form $\boldsymbol{\omega}\in T^{*}_x\mathcal{S}\,$, their natural pairing is denoted by $\langle \boldsymbol{\omega},\mathbf{u} \rangle=\boldsymbol{\omega}(\mathbf{u})=\omega_a\,\mathrm{u}^a\,$.
Similarly, in the reference manifold $\mathcal B\,$, given a vector $\mathbf{U} \in T_X\mathcal{B}$ and a 1-form $\boldsymbol{\Omega} \in T^*_X\mathcal{B}\,$, their natural pairing is also denoted by $\langle \boldsymbol{\Omega}, \mathbf{U} \rangle = \boldsymbol{\Omega}(\mathbf{U}) = \Omega_A \mathrm{U}^A\,$. Note that the natural pairing operation does not require any metric structure.

As a measure of strain in elastic solids, we typically use the derivative of the deformation mapping\textemdash known as the deformation gradient\textemdash denoted by $\F(X,t)=T\varphi_t(X):T_X\mathcal{B}\to T_{\varphi_t(X)}\mathcal{C}_t$; in components it reads as ${\cF^a}_A=\frac{\partial \varphi^a}{\partial X^A}\,$.
The adjoint $\F^\star$ of $\F$ is defined as ${\F^\star(X,t):T_{\varphi_t(X)}\mathcal{C}_t \to T_X\mathcal{B}}\,$,  ${\langle\boldsymbol\alpha,\F \mathbf{U}\rangle=\langle\F^\star \boldsymbol\alpha,\mathbf{U}\rangle}\,$,  $\forall\, \mathbf{U} \in T_X\mathcal B\,$,  $\forall\, \boldsymbol\alpha \in T^{*}_{\varphi(X)}\mathcal S$; it has components ${\left(\cF^\star\right)_A}^a={\cF^a}_A\,$. Note that the definition of the adjoint, much like the natural pairing, does not require any metric structure either.
 
The ambient space has a background Euclidean metric $\g=\mathrm{g}_{ab}\,dx^a\otimes dx^b\,$. Given vectors $\mathbf{u}\,,\mathbf{w}\in T_x\mathcal{S}\,$, their dot product is denoted by $\llangle \mathbf{u},\mathbf{w} \rrangle_{\g}=\mathrm{u}^a\,\mathrm{w}^b\,\mathrm{g}_{ab}\,$. The spatial volume form is $dv = \sqrt{\det\g}\,\,dx^1 \wedge dx^2 \wedge dx^3\,$. The Levi-Civita connection of $(\mathcal{S},\g)$ is denoted by $\bar\nabla\,$, with Christoffel symbols ${\gamma^a}_{bc}\,$.

\begin{remark}\label{rmrk:g_metric}
The same notation $\mathbf{g}$ is used for both the background metric of the ambient space $\mathcal{S} = \mathbb{R}^3$ and the spatial metric of the deformed configuration $\varphi_t(\mathcal{B})\,$, but they have a subtle yet important difference. The background metric is fixed and time-independent, describing the intrinsic geometry of the ambient space. In contrast, the spatial metric evolves over time due to its dependence on the deformation mapping $\varphi_t\,$, as its domain of definition is the deformed configuration $\varphi_t(\mathcal{B})\,$.
Formally, the spatial metric $\mathbf g_t$ is the pull-back of the background metric $\mathbf g$ by the time-dependent inclusion map $\iota_t:\varphi_t(\mathcal B)\to\mathcal S\,$, i.e.~$\mathbf g_t=\iota_t^*\mathbf{g}\,$.
This distinction is essential: the background metric provides a constant geometric reference for measuring local distances, whereas the spatial metric reflects the dynamic geometry of the deforming body. For simplicity, we use the notation $\mathbf{g}$ for the spatial metric in the rest of the paper.
\end{remark}

The material velocity of the motion is defined as ${\mathbf{V}:\mathcal{B}\times\mathbb{R}^+\to T \mathcal{S}\,, \mathbf{V}(X,t)\coloneq \frac{\partial \varphi(X,t)}{\partial t}}$; it has components $\mathrm{V}^a = \frac{\partial \varphi^a}{\partial t}\,$.
The spatial velocity is defined as ${\mathbf{v}:\varphi_t(\mathcal{B})\times\mathbb{R}^+\to T \mathcal{S}}\,$,  ${\mathbf{v}(x,t)\coloneq \mathbf{V}(\varphi_t^{-1}(x),t)}\,$.
The material acceleration of the motion is defined as ${\mathbf{A}:\mathcal{B}\times\mathbb{R}^+\to T \mathcal{S}}\,$,  ${\mathbf{A}(X,t)\coloneq D_t^{\g}\mathbf{V}(X,t)}\,$,  where $D_t^{\g}$ denotes the covariant derivative along $\varphi_X:t\mapsto \varphi(X,t)$; it reads in components as $\mathrm{A}^a=\frac{\partial \mathrm{V}^a}{\partial t}+{\gamma^a}_{bc}\,\mathrm{V}^b\,\mathrm{V}^c\,$.
The spatial acceleration of the motion is defined as $\mathbf{a}:\varphi_t(\mathcal{B})\times\mathbb{R}^+\to T \mathcal{S}\,$,  $\mathbf{a}(x,t)\coloneq \mathbf{A}(\varphi_t^{-1}(x),t)\in T_x\mathcal{S}$; it has components $\mathrm{a}^a=\frac{\partial\mathrm{v}^a}{\partial t}+\frac{\partial\mathrm{v}^a}{\partial x^b}\,\mathrm{v}^b+{\gamma^a}_{bc}\,\mathrm{v}^b\,\mathrm{v}^c\,$.

\subsection{The Material Configuration in Finite Elasticity}
\label{Sec:EuclidG}

In the context of finite elasticity, the body’s natural reference configuration, also referred to as its material configuration, is stress-free and at rest within the Euclidean ambient space. Specifically, $\mathcal{B} \subset \mathcal{S} = \mathbb{R}^3$ and naturally inherits its geometry from the ambient space, meaning its metric is essentially a copy of the background metric $\mathbf{g}\,$. Denoting the material metric in finite elasticity by $\Go\,$, it is formally defined as the pullback of the background metric under the inclusion embedding $\iota: \mathcal{B} \to \mathcal{S}\,$, such that $\Go = \iota^* \mathbf{g}\,$.

The material (natural stress-free) configuration for elasticity is $(\mathcal{B}, \Go)$ where distances and angles are measured using the material metric $\Go\,$, which serves as the baseline against which deformations are quantified.\footnote{Refering to Remark~\ref{rmrk:g_metric}, note that, at time $t=t_0\,$, before the deformation begins, the deformation mapping reduces to the identity map on $\mathcal B\,$, i.e.~$\varphi_{t_0}=\operatorname{id}_{\mathcal{B}}\,$, and the spatial metric at $t=t_0$ reduces to the Euclidean material metric, i.e. $\g_{t_0}=\Go\,$.}
Given two referential vectors $\mathbf{U}, \mathbf{W} \in T_X \mathcal{B}\,$, their dot product in the material configuration $(\mathcal{B}, \Go)$ is expressed as $\llangle \mathbf{U}, \mathbf{W} \rrangle_{\Go} = \mathrm{U}^A \mathrm{W}^B \mathring{\mathrm{G}}_{AB}\,$.
The material volume form given by $d\mathring{V} = \sqrt{\det \Go} \, dX^1 \wedge dX^2 \wedge dX^3\,$. The Levi-Civita connection associated with $(\mathcal{B}, \Go)\,$, denoted by $\mathring\nabla\,$, has Christoffel symbols ${\mathring\Gamma^A}_{BC}\,$.
The motion $\varphi: \mathcal{B} \to \mathcal{S}$ maps each material point in the reference configuration to its spatial position in the deformed configuration, and its Jacobian $\mathring{J}$ relates the referential and spatial volume elements as $\varphi^* dv = \mathring{J} d\mathring{V}\,$. It can be shown that $\mathring{J} = \sqrt{\det\g/\det\Go} \det\F\,$.

\subsection{The Material Configuration in Finite Anelasticity}
\label{Sec:NonEuclidG}

Anelasticity describes a class of material behaviours where intrinsic distortions, referred to as eigenstrains\textemdash derived from the German word \textit{Eigenspannungsquellen} (sources of inherent stresses) introduced by Reissner \citep{Reissner1931EigenspannungenUE}, contribute to the overall deformation of an otherwise elastic body. These eigenstrains, arising from non-elastic deformations or microstructural changes such as plasticity, growth, swelling, or thermal expansion, lead to a residually-stressed state when the body is relaxed in the Euclidean ambient space. Indeed, due to the presence of eigenstrains, the natural (stress-free) configuration of the body generally cannot be isometrically embedded in the Euclidean ambient space, meaning it cannot exist as a stress-free body in the physical space. The resulting residual stresses are, therefore, an expression of the non-Euclidean nature of the natural configuration. Consequently, the manifold $\mathcal{B}$ must be endowed with a material metric $\G$ that is generally distinct from the Euclidean metric $\Go\,$.

To model the interplay between intrinsic, often irreversible and dissipative, distortions (eigenstrains) and the recoverable, nondissipative elastic deformations in anelastic materials, we introduce the multiplicative decomposition of the total deformation gradient: $\F = \Fe \Fa$\textemdash known as the Bilby-Köner-Lee decomposition \citep{lubarda2002multiplicative, Sadik2017}. Here, $\Fa$ maps the Euclidean reference configuration $(\mathcal{B}, \Go)$ to the stress-free, but generally non-Euclidean, material configuration for anelasticity $(\mathcal{B}, \mathbf{G})\,$, while $\Fe$ represents the recoverable elastic deformation from the material (stress-free) configuration to the current configuration.
Note, however, that while the total deformation gradient $\mathbf{F}$ is compatible\textemdash meaning it can be expressed as the derivative of a smooth mapping $\varphi\,$, such that $\mathbf{F} = T\varphi$\textemdash the individual distortions $\Fe$ and $\Fa$ are generally incompatible. This incompatibility arises because $\Fe$ and $\Fa$ are not derivatives of smooth mappings themselves but instead represent local, often non-homogeneous, distortions.
As a matter of fact, these residual incompatibilities in $\Fe$ and $\Fa$ are a direct manifestation of the failure of the relaxed configuration to be embedded in the physical Euclidean space, and effectively justify the non-Euclidean character of the material configuration preventing the fully relaxed state from being realized as a smooth, globally compatible embedding.
The material metric $\mathbf{G}$ associated with the stress-free configuration is related to the Euclidean reference metric $\Go$ through the relationship $\mathbf{G} = \Fa^* \Go\,$.\footnote{To see this, consider a curve in the reference configuration, $\gamma: \mathfrak{I} \to \mathcal{B}\,$, where $\mathfrak{I}$ is an open interval in $\mathbb{R}\,$. Due to the presence of eigenstrains, the curve $\gamma$ is generally stressed in the Euclidean reference configuration $(\mathcal{B}, \mathring{\mathbf{G}})\,$. However, its push-forward under the anelastic distortion, $\Fa_* \gamma\,$, is locally stress-free. The squared arc-length element in this natural (stress-free) configuration is given by $\llangle \Fa_* \gamma'(t), \Fa_* \gamma'(t) \rrangle_{\mathring{\mathbf{G}}} = \llangle \Fa \gamma'(t), \Fa \gamma'(t) \rrangle_{\mathring{\mathbf{G}}} = \llangle \gamma'(t), \gamma'(t) \rrangle_{\Fa^* \mathring{\mathbf{G}}}$\textemdash see \citep{Yavari2021}. Hence, the material metric $\mathbf{G}\,$, which measures natural distances in the stress-free configuration, is the pull-back of the Euclidean metric $\mathring{\mathbf{G}}$ under the anelastic distortion $\Fa\,$.}

Note that multiple sources of eigenstrains can coexist within a solid, including temperature changes, growth, remodeling, swelling, and plastic deformation. Each eigenstrain distribution contributes a distinct distortion field, leading to a combined total anelastic distortion. This total distortion can be expressed as a product of individual contributions: $\Fa = \prod_{j=1}^{N} \accentset{j}{\mathbf{F}} = \accentset{1}{\mathbf{F}} \hdots \accentset{N}{\mathbf{F}}\,$, where $\accentset{j}{\mathbf{F}}$ represents the distortion field associated with the $j$-th anelastic process.

\begin{remark}[Thermal Distortion in Anelasticity]
In the case of a non-isothermal process, characterised by an evolving temperature field $\Theta = \Theta(X, t)$, thermal distortion emerges as a distinct source of anelastic distortion. This distortion captures the local, generally incompatible, stretches induced by temperature variations \citep{stojanovic1964finite, Sadik2017Thermoelasticity}.
The total anelastic distortion $\Fa$ can hence be decomposed as $\Fa = \Fn \Fth\,$, where $\Fth$ accounts for thermal distortion, while $\Fn$ encompasses all other anelastic mechanisms coexisting in the solid.
What distinguishes thermal distortion is that its evolution is governed by the heat equation, making it directly dependent on the temperature field $\Theta = \Theta(X, t)\,$. We show in Appendix~\ref{App:Thermal} that
\begin{equation}
\label{eq:Th_distort}
\boldsymbol\alpha = \frac{\partial\Fth}{\partial\Theta} \,\Fth^{-1} = \Fth^{-1} \frac{\partial\Fth}{\partial\Theta}\,,
\end{equation}
where $\boldsymbol\alpha$ is the thermal expansion coefficient tensor of the material.
\end{remark}

For an anelastic solid, the material (natural stress-free) configuration is hence given by the abstract manifold $(\mathcal{B}, \mathbf{G})$ where distances and angles are measured using  the Riemannian metric $\G\,$. Given two vectors $\mathbf{U}, \mathbf{W} \in T_X\mathcal{B}\,$, their dot product is expressed as: $\llangle \mathbf{U}, \mathbf{W} \rrangle_{\mathbf{G}} = \mathrm{U}^A \mathrm{W}^B \mathrm{G}_{AB}\,$. The volume form in the anelastic material configuration is given by $dV = \sqrt{\det \mathbf{G}} \, dX^1 \wedge dX^2 \wedge dX^3\,$. The Levi-Civita connection associated with $(\mathcal{B}, \mathbf{G})$ is denoted $\nabla$ with Christoffel symbols ${\Gamma^A}_{BC}\,$.
In anelasticity, the elastic Jacobian $\Je$ of the motion relates the stress-free material and spatial volume elements as $\varphi^* dv = \Je dV\,$, and it can be shown that $\Je = \sqrt{\det\g/\det\G} \det\F = \sqrt{\det\g/\det\Go}\, \det\Fe\,$. We may also introduce the athermal anelastic Jacobian related to the athermal anelastic distortion $\Fn$ as $\Jn = \det\Fn\,$.

\section{Thermodynamics and the Balance Laws of Hyperelasticity}
\label{Sec:ThermoElast}

Hyperelasticity is a subclass of elasticity in which the stress is derived from a scalar strain energy function as discussed below in Remark~\ref{rmrk:hyper}. Recall that in finite elasticity, the material metric $\Go$ is a copy of the background Euclidean metric $\g\,$, i.e. $\Go = \iota^*\g\,$, and serves as the geometric foundation for defining distances and angles in the stress-free reference configuration\textemdash see \S\ref{Sec:EuclidG}.
In what follws, we begin by briefly reviewing the first and second laws of thermodynamics in the setting of nonlinear hyperelasticity. We then demonstrate, by introducing an extension to the concept of \citet{coleman1963}'s thermodynamic process, how all the balance laws and the constitutive equations of hyperelasticity can be derived directly from these thermodynamic principles, without invoking (observer) invariance.\footnote{As discussed earlier in \S\ref{S:intro}, the Green-Naghdi-Rivlin theorem (and its subsequent extensions) gives the balance of linear and angular momenta as a consequence of the invariance of the energy balance under superposed isometries of the ambient space. Here, the proposed derivation does not rely on invariance; instead, it generalises the Coleman-Noll procedure \citep{coleman1963} to recover not only the constitutive equations, but also the balance laws.}

\begin{remark}[Hyperelasticity]
\label{rmrk:hyper}
As formalized by Noll \citep{noll1958}, a \textit{simple elastic solid} is that for which the stress at a given point depends only on the local current deformation state at that point (e.g., via the Finger tensor $\mathbf{b}$), discarding any dependence on the material’s deformation history or nonlocal effects such as higher-order spatial gradients.
A particular subclass of simple elastic materials is known as \textit{Cauchy elastic} materials, where the stress at any point is explicitly expressed as a function of the strain at that point, such as $\boldsymbol{\sigma} = \boldsymbol{\mathsf{f}}(\mathbf{b})$ \citep{Cauchy1828, Truesdell1952, Yavari2024Cauchy}. This subclass assumes a direct and explicit dependence of stress on strain, further refining the constitutive description.
Within Cauchy elasticity, another important subset consists of materials for which the stress is derivable from a scalar strain energy function. These are referred to as \textit{Green elastic} \citep{Green1838, Green1839, Spencer2015} or \textit{hyperelastic} \citep{Truesdell1952}. However, not all Cauchy elastic materials possess such an energy function, as Green elasticity imposes additional constraints on the constitutive relation.
It is worth noting that not all simple elastic solids are Cauchy elastic. There are those whose constitutive relations are defined implicitly, taking the form $\boldsymbol{\mathsf{f}}(\boldsymbol{\sigma}, \mathbf{b}) = \mathbf{0}\,$, where the stress and deformation measures are related implicitly \citep{Morgan1966, Rajagopal2003, Rajagopal2007, Rajagopal2011,Bustamante2009,Bustamante2011,Yavari2024ImplicitElasticity}. These implicit models encompass both Cauchy elastic materials and other, more complex elastic behaviours. Thus, Cauchy elastic solids form a proper subset of simple elastic materials, and Green elastic materials are a further subset of Cauchy elastic solids. This hierarchy reflects the progressively restrictive assumptions underpinning these material models.
\end{remark}

\subsection{The First Law of Thermodynamics}
\label{Sec:1stLaw-Elast}

The first law of thermodynamics posits the existence of an internal energy $\mathscr{E}$ as a state function, which satisfies the following balance equation as an expression of the conservation of energy principle \citep{Truesdell1952, gurtin1974modern, MarsdenHughes1983}
\begin{equation}\label{eq:Thermo_First}
	\frac{d}{dt}
	\int_{\mathcal{U}} \mathring\rho \left(\mathscr{E}+\frac{1}{2} \Vert\mathbf V\Vert^2_{\g}\right)
	d\mathring{V}
	=\int_{\mathcal{U}}\mathring\rho \left(\llangle\mathbf{B},\mathbf{V}\rrangle_{\g}
	+R\right)d\mathring{V}+\int_{\partial \mathcal{U}}
	\left(\llangle\mathbf{T},\mathbf{V}\rrangle_{\g}-\llangle \mathbf{Q},\mathbf{N}\rrangle_{\Go}\right)d\mathring{A}\,,
\end{equation}
for any open set $\mathcal{U}\subset\mathcal{B}\,$, where $\mathscr{E}$ stands for the specific internal energy, $\mathring\rho$ is the material mass density, $\mathbf{B}$ is the specific body force, ${R=R(X,t)}$ is the specific heat supply, $\mathbf{T}=\mathbf{T}(X,\mathbf{N})$ is the boundary traction vector field per unit material area, $\mathbf{N}$ is the $\Go$-unit normal to the boundary $\partial\mathcal U\,$, and ${\mathbf Q=\mathbf Q(X,\Theta,d\Theta,\F,\Go,\g)}$ represents the external heat flux per unit material area, $d\Theta=\frac{\partial \Theta}{\partial X^A}dX^A$ is the exterior derivative of temperature $\Theta\,$, and $d\mathring{A}$ is the material area element.

Using Marsden and Hughes's version of Cauchy's theorem \citep[\S2.1(1.9)]{MarsdenHughes1983},\footnote{Note that this version of Cauchy's theorem does not assume the balance of linear momentum and does not introduce the idea of stress\textemdash and neither do we, at this point in the paper.} it follows from \eqref{eq:Thermo_First} that there exists a unique material vector field $\mathbf{U}$ such that $\llangle \mathbf{U},\mathbf{N}\rrangle_{\Go}=\llangle\mathbf{T},\mathbf{V}\rrangle_{\g}\,$. The linearity of $\llangle\mathbf{T},\mathbf{V}\rrangle_{\g}$ with respect to $\mathbf{V}$ implies that $\mathbf{U}$ is also linear in $\mathbf{V}\,$, i.e. there exists a second-order two-point tensor field $\mathbf{M}$ such that $\mathbf{U}=\mathbf{M} \mathbf{V}\,$, which indeed is unique following the uniqueness of $\mathbf{U}\,$. We now have $\llangle \mathbf{M} \mathbf{V},\mathbf{N}\rrangle_{\Go}=\llangle\mathbf{T},\mathbf{V}\rrangle_{\g}\,$, which may be recast into $\llangle \mathbf{M}^{\mathring{\mathsf T}} \mathbf{N},\mathbf{V}\rrangle_{\g}=\llangle\mathbf{T},\mathbf{V}\rrangle_{\g}\,$. By virtue of the existence and uniqueness of $\mathbf{M}\,$, we may define the two-point tensor $\mathbf{P}:=\mathbf{M}^{\mathring{\mathsf T}}\g^\sharpo\,$, and it follows by arbitrariness of $\mathbf{V}$ that $\mathbf{T}=\mathbf{P} \mathbf{N}^\flato\,$, where $(.)^\flato$ and $(.)^\sharpo$ denote the musical isomorphisms for lowering and raising indices, respectively, with respect to $\Go$ and $\g\,$.
Now, we proceed to write the energy balance \eqref{eq:Thermo_First} in local form:
\begin{equation}\label{First-Law-Local}
	\mathring\rho\,\dot{\mathscr{E}} =
	\mathbf P\!:\! (\g \bar\nabla \mathbf V)
	- \operatorname{\mathring{D}iv} \mathbf Q +\mathring\rho R
	+ \llangle \operatorname{\mathring{D}iv}\mathbf{P}+ \rho(\mathbf{B}
	- \mathbf{A}),\mathbf{V} \rrangle_{\g}
	-\dot{\rho}_0 \left(\mathscr{E}+\frac{1}{2} \Vert\mathbf V\Vert^2_{\g}\right) \,,
\end{equation}
where a dotted quantity denotes its total time derivative, $\operatorname{\mathring{D}iv}$ denotes the two-point divergence operator with respect to the connections $\mathring\nabla$ and $\bar\nabla\,$, and a colon denotes the double contraction product, i.e. $\mathbf P\!:\! (\g \bar\nabla \mathbf V) = \textrm{P}^{aA}\textrm{V}_{a|A}$\textemdash the vertical stroke here denoting covariant differentiation with respect to $\g$ in components, i.e. $\textrm{V}_{a|A} = \textrm{V}_{a,A} - \textrm{V}_b \gamma^b_{ac} {\textrm{F}^c}_{A}$. 
Note that\footnote{We thank Sanjay Govindjee for bringing to our attention that $\g \bar\nabla \mathbf V$ is not necessarily symmetric.}
\begin{equation} \label{eq:rate-decomp}
	\mathbf P\!:\! (\g \bar\nabla \mathbf V) 
	= \mathbf P\!:\! (\g \bar{\operatorname{D}}_t \F) = (\F^{-1} \mathbf P) \!:\! (\Ds+ \Da)\,,
\end{equation}
where $\bar{\operatorname{D}}_t (.)$ denotes the covariant time derivative in $(\mathcal S,\mathbf g)\,$,\footnote{In components $\bar{\operatorname{D}}_t {\cF^a}_A=\frac{\partial }{\partial t}\!\left({\cF^a}_A\right) + {\cF^b}_A {\gamma^a}_{bc} \mathrm{v}^c\,$. Note that by using the symmetry lemma \citep{Nishikawa2002}, one may write ${V^a}_{|A}=\left(\frac{\partial \varphi^a}{\partial t}\right){\!}_{|A} = \bar{\operatorname{D}}_t\left(\frac{\partial \varphi^a}{\partial X^A}\right)\,$, which implies that $\bar\nabla \mathbf V = \bar{\operatorname{D}}_t\F\,$.}
$\Ds$ is the symmetric part of $\F^\star \g \bar{\operatorname{D}}_t \F\,$, and $\Da$ is its anti-symmetric part:
\begin{equation}
\label{Eq:Def_rates}
	\Ds
	= \frac{1}{2}\left[\F^\star \g \bar{\operatorname{D}}_t \F
	+ \bar{\operatorname{D}}_t \F^\star \g \F\right]\,,
	\qquad
	\Da
	= \frac{1}{2}\left[\F^\star \g \bar{\operatorname{D}}_t \F
	- \bar{\operatorname{D}}_t \F^\star \g \F\right]
	 \,.
\end{equation}

\subsection{The Second Law of Thermodynamics}

The second law of thermodynamics posits the existence of specific entropy $\mathscr{N}$ as a state function, which satisfies the following inequality\textemdash known as the Clausius-Duhem inequality\textemdash as an expression of the principle of entropy production,\footnote{The entropy production for an open subset $\mathcal U$ in the body is written as $$\Gamma=\frac{d}{dt} \int_{\mathcal{U}} \mathring\rho \mathscr{N}d\mathring{V}-\int_{\mathcal{U}}\mathring\rho \frac{R}{\Theta}d\mathring{V}-\int_{\partial\mathcal{U}}\frac{H}{\Theta}\,d\mathring{A}\,.$$} which steadily increases or remains constant within a closed system over time \citep{Truesdell1952,gurtin1974modern,MarsdenHughes1983}
\begin{equation}\label{eq:Thermo_Second}
	\frac{d}{dt} \int_{\mathcal{U}} \mathring\rho \mathscr{N}d\mathring{V}
	\geq\int_{\mathcal{U}}\mathring\rho \frac{R}{\Theta}d\mathring{V}-\int_{\partial\mathcal{U}}\frac{1}{\Theta}\llangle \mathbf{Q},\mathbf{N}\rrangle_{\Go}d\mathring{A}\,,
\end{equation}
for any open set $\mathcal{U}\subset\mathcal{B}\,$.
In localized form, the material Clausius-Duhem inequality \eqref{eq:Thermo_Second} is written as
\begin{equation} \label{Second-Law-Local}
	\dot\eta = \mathring\rho \Theta \dot{\mathscr{N}}+\dot{\mathring\rho} \Theta \mathscr{N} 
	+ \operatorname{\mathring{D}iv}\mathbf{Q}  - \mathring\rho R -\frac{1}{\Theta}\langle d\Theta,\mathbf{Q}\rangle
	\geq 0\,,
\end{equation}
where $\dot\eta$ denotes the material rate of energy dissipation density.

\begin{remark}
\label{rmrk:thermodynamic_process}
At a material point $X\in\mathcal{B}\,$, \citet{coleman1963} introduced the set\footnote{$\mathbf{P}\,$, in their treatment, is introduced, a priori, as the first Piola-Kirchhoff stress tensor (in fact, they use the Cauchy stress tensor) satisfying the balance of linear and angular momenta.}
\begin{equation} \label{Dynamical-Process}
	\Big\{
	\varphi_t(X),\mathbf{P}(X,t),\mathbf{B}(X,t),\mathscr{E}(X,t),\mathbf{Q}(X,t),
	R(X,t),\mathscr{N}(X,t),\Theta(X,t)
	\Big\}	
	\,,
\end{equation}
and called it a \emph{thermodynamic process}, provided all its eight fields satisfy the first law of thermodynamics and the balance of linear and angular momenta.
A given hyperelastic material is specified by its constitutive assumptions, e.g., $\mathscr{E}=\mathscr{E}(X,\mathscr{N},\F,\Go,\g)$ and ${\mathbf Q=\mathbf Q(X,\Theta,d\Theta,\F,\Go,\g)}\,$. A thermodynamic process is \emph{admissible} if the constitutive assumptions hold everywhere in the body and at all times.
\citet{coleman1963} showed that requiring the Clausius-Duhem inequality \eqref{Second-Law-Local} to hold for all admissible thermodynamic processes puts certain constraints on the constitutive assumptions, e.g., the Doyle-Ericksen formula.
\end{remark}


\begin{defi}\label{def:Extended}
An \emph{extended thermodynamic process} at a material point $X\in\mathcal{B}$ is defined as the set
\begin{equation} \label{Ext-Dynamical-Process}
	\left\{
	\mathring{\rho}(X,t), \varphi_t(X), \mathbf{P}(X,t), \mathbf{B}(X,t), \mathscr{E}(X,t), 
	\mathbf{Q}(X,t), R(X,t), \mathscr{N}(X,t), \Theta(X,t) \right\}	\,,
\end{equation}
provided all its nine fields satisfy the first law of thermodynamics\textemdash without requiring that $\left\{\mathring{\rho}(X,t),\varphi_t(X),\mathbf{P}(X,t),\mathbf{B}(X,t)\right\}$ satisfy the balance of mass, linear, or angular momenta. An extended thermodynamic process is \emph{admissible} if the constitutive assumptions for
$\left\{\mathscr{E}(X,t),\mathbf{Q}(X,t), R(X,t),\mathscr{N}(X,t)\right\}$
hold everywhere in the body and at all times.
We require that the Clausius-Duhem inequality \eqref{Second-Law-Local} holds for all admissible extended thermodynamic processes.
\end{defi}

\begin{remark}
\label{remark:RET1}
This definition marks a key departure from the classical Coleman–Noll procedure: the balance of mass and momenta are not imposed a priori. 
Instead, we prove in what follows that these balance laws emerge as necessary conditions for the admissibility of an extended process. This shift allows us to treat the entropy inequality as a generative principle for both constitutive relations and balance laws. 
The underlying perspective is conceptually aligned with Rational Extended Thermodynamics (RET), developed by M\"uller and Ruggeri \citep{Muller1972,MullerRuggeri1998RET,Ruggeri2012a,Ruggeri2012b,Ruggeri2024}, where fundamental principles are also used to generate, rather than merely constrain, field equations. 
However, as we elaborate in Remark~\ref{remark:RET2}, our notion of an extended process differs fundamentally from what is meant by extension in RET: here, the extension pertains to including the mass density and the a priori conditions a thermodynamic processes ought to satisfy, whereas in RET, it involves an enlargement of the state space by introducing additional higher-order variables as independent fields.
\end{remark}

\subsection{Constitutive Equations and Balance Laws in Hyperelasticity}

Hyperelasticity implies the existence an energy function that depends explicitly at every material point $X \in \mathcal{B}$ on the strain at that same point.
We may hence write the specific free energy as $\Psi = {\Psi}(X,\Theta,\F,\Go,\g)$ \citep{Truesdell1952}, which in fact is the Legendre transform of the specific internal energy $\mathscr E$ with respect to the conjugate variables temperature $\Theta$ and specific entropy $\mathscr N$:
\begin{equation}\label{Free-Energy}
	\Psi = \mathscr E - \Theta \mathscr N\,.
\end{equation}
Hence, we have that ${\mathscr E = {\mathscr E}(X, \mathscr N,\F,\Go,\g)}\,$, and consequently have\footnote{Note that the Legendre transform \eqref{Free-Energy} of $\mathscr{E}$ to $\Psi$ with respect to the conjugate variables $\mathscr{N}$ and $\Theta$ is essentially a change of variable satisfying \eqref{Entropy-Temperature}. See \citep{Arnold1989ClassMech,Goldstein2002ClassMech} for further details on Legendre transform in the context of Lagrangian mechanics and thermodynamics.}
\begin{equation} \label{Entropy-Temperature}
	\mathscr N = -\frac{\partial \Psi}{\partial \Theta}\,.
\end{equation}

\begin{prop}\label{prop:hyper}
For a hyperelastic body, the first and second laws of thermodynamics \eqref{First-Law-Local} and \eqref{Second-Law-Local} imply that
\begin{subequations}\label{eq:prop_hyper}
\begin{empheq}[left={\empheqlbrace}]{align}
	\label{Mass-Conservation}
	& \dot{\mathring\rho}=0 \,, \\
	\label{Doyle-Ericksen}
	& \mathbf P=\mathring\rho\g^\sharpo \frac{\partial \Psi}{\partial\F}\,,\\
	\label{Linear-Momentum}
	& \operatorname{\mathring{D}iv}\mathbf{P}+ \mathring\rho \mathbf{B} =\mathring\rho\mathbf{A} \,, \\
	\label{Angular-Momentum}
	& \F\mathbf{P}^{\star}=\mathbf{P}\F^{^\star}\,,\\
	\label{Diss_Ineq}
	& \dot\eta = - \frac{1}{\Theta} \langle d\Theta, \mathbf Q \rangle 	
	\geq 0\,.
\end{empheq}
\end{subequations}
In other words, the first and second laws of thermodynamics imply the conservation of mass \eqref{Mass-Conservation}, the Doyle-Ericksen formula \eqref{Doyle-Ericksen}, the balance of linear momentum \eqref{Linear-Momentum}, the balance of angular momentum \eqref{Angular-Momentum}, and the dissipation inequality \eqref{Diss_Ineq}. Note that \eqref{Doyle-Ericksen} effectively shows that the two-point tensor $\mathbf{P}$ is indeed the first Piola-Kirchhoff stress tensor.
\end{prop}
\begin{proof}
From \eqref{Free-Energy}, one writes: $\mathring\rho \Theta \dot{\mathscr{N}}=\mathring\rho \dot{\mathscr{E}}-\mathring\rho\dot{\Psi}-\mathring\rho \dot{\Theta} \mathscr{N}\,$. Substituting this relation and \eqref{First-Law-Local} (while using \eqref{eq:rate-decomp}) into \eqref{Second-Law-Local}, one obtains
\begin{equation} \label{Dissipation1}
\begin{split}
	\dot\eta =& (\F^{-1} \mathbf P)\!:\! (\Ds+\Da)-\mathring\rho\dot{\Psi}-\mathring\rho \dot{\Theta} \mathscr{N}
	+\dot{\rho}_0 \Theta \mathscr{N}\\
	&+ \llangle \operatorname{\mathring{D}iv}\mathbf{P}+ \rho(\mathbf{B} - \mathbf{A}),\mathbf{V} \rrangle_{\g}
	-\dot{\rho}_0 \left[\mathscr{E}+\frac{1}{2} \Vert\mathbf V\Vert^2_{\g}\right]
	-\frac{1}{\Theta}\langle d\Theta,\mathbf{Q}\rangle
	\geq 0\,.
\end{split}
\end{equation}
Applying Leibniz rule to $\dot\Psi\,$, one writes\footnote{Note that since the connection is Levi-Civita, which is metric compatible, it follows that $\bar{\operatorname{D}}_t \g=\bar\nabla_{\mathbf v}\g = \mathbf 0\,$.}
\begin{equation}\label{eq:psi-dot}
	\dot{\Psi}=
	\frac{\partial \Psi}{\partial \Theta}\dot{\Theta}
	+\frac{\partial \Psi}{\partial \F}\!:\!\bar{\operatorname{D}}_t \F
	+\frac{\partial \Psi}{\partial \g}\!:\!\bar{\operatorname{D}}_t \g
	=-\mathscr N \dot{\Theta}
	+ \left(\F^{-1} \g^\sharpo \frac{\partial \Psi}{\partial \F} \right)\!:\! (\Da +\Ds)
	\,.
\end{equation}
It can be shown that\footnote{Using Leibniz rule, we write in components
\begin{equation} \nonumber
	\frac{\partial \Psi}{\partial \textrm{F}^a{}_A}
	=\frac{\partial \Psi}{\partial (\textrm{F}^i{}_K \,\textrm{F}^j{}_L \,\textrm{g}_{ij})}
	\frac{\partial (\textrm{F}^k{}_K \,\textrm{F}^l{}_L \,\textrm{g}_{kl})}{\partial \textrm{F}^a{}_A}
	=\frac{\partial \Psi}{\partial (\textrm{F}^i{}_K \,\textrm{F}^j{}_L \,\textrm{g}_{ij})}
	\left(
	\delta^k{}_a \delta^A{}_K \,\textrm{F}^l{}_L \,\textrm{g}_{kl}
	+ \textrm{F}^k{}_K \,\delta^l{}_a \delta^A{}_L \,\textrm{g}_{kl}
	\right)
	\,.
\end{equation}
Using straightforward index manipulations, it follows that
\begin{equation} \nonumber
	\frac{\partial \Psi}{\partial \textrm{F}^a{}_A}
	= 2 \textrm{g}_{al}\, \textrm{F}^l{}_K \,
	\frac{\partial \Psi}{\partial (\textrm{F}^i{}_K \,\textrm{F}^j{}_A \,\textrm{g}_{ij})}\,,  
	\quad\textrm{and}\qquad
	\frac{\partial \Psi}{\partial \textrm{F}^a{}_A}
	=2\frac{\partial \Psi}{\partial (\textrm{F}^i{}_A \textrm{F}^j{}_L \textrm{g}_{ij})}
	\,  \textrm{F}^k{}_L \,\textrm{g}_{ka}\,,
\end{equation}
which one may rewrite as
\begin{equation} \nonumber
	\frac{\partial \Psi}{\partial \F}
	= 2 \g \F \frac{\partial \Psi}{\partial (\F^{\star} \g \F)}\,,
	\quad\textrm{and}\qquad
	\left(\frac{\partial \Psi}{\partial \F}\right)^\star
	= 2 \frac{\partial \Psi}{\partial (\F^{\star} \g \F)} \F^\star \g\,.
\end{equation}
Therefore, one finds that
\begin{equation} \nonumber
	\F^{-1} \g^\sharpo \frac{\partial \Psi}{\partial \F}
	=
	\left(\frac{\partial \Psi}{\partial \F}\right)^\star\!\g^\sharpo \F^{-\star}\,,
\end{equation}
which directly leads to \eqref{eq:F-symmetry}.
}
\begin{equation} \label{eq:F-symmetry}
	\left(\F^{-1} \g^\sharpo \frac{\partial \Psi}{\partial \F}\right)^\star = \F^{-1} \g^\sharpo \frac{\partial \Psi}{\partial \F}
	\,.
\end{equation}
Hence, $\F^{-1} \g^\sharpo \frac{\partial \Psi}{\partial \F}$ is a symmetric tensor, and since $\Da$ is antisymmetric, it follows that
\begin{equation}
\left(\F^{-1} \g^\sharpo \frac{\partial \Psi}{\partial \F} \right)\!:\! \Da = 0
	\,.
\end{equation}
Using the above identity, and substituting \eqref{eq:psi-dot} into \eqref{Dissipation1}, the rate of dissipation is simplified to read
\begin{equation}\label{Dissipation1b}
\begin{split}
	\dot\eta &= \left[\F^{-1} \left(\mathbf{P} - \mathring\rho \g^\sharpo \frac{\partial \Psi}{\partial \F}\right)\right]\!\!:\! \Ds
	+ \F^{-1}\mathbf{P}\!:\! \Da
	\\&\quad
	+ \llangle \operatorname{\mathring{D}iv}\mathbf{P}+ \rho(\mathbf{B} 
	- \mathbf{A}),\mathbf{V} \rrangle_{\g}
	-\dot{\rho}_0 \left[\Psi+\frac{1}{2} \Vert\mathbf V\Vert^2_{\g}\right]
	-\frac{1}{\Theta}\langle d\Theta,\mathbf{Q}\rangle
	\geq 0\,.
\end{split}
\end{equation}
This inequality must hold for all motions, i.e. all extended thermodynamic processes. 
As $\Ds$ (a symmetric tensor) and $\Da$ (an antisymmetric tensor) can be chosen independently of all the other fields, one concludes that\footnote{There are infinitely may extended thermodynamic processes for which everything except for $\Ds$ and $\Da$ are the same. For the second law to hold for all such processes one must have \eqref{DE-AngMom}.}
\begin{equation} \label{DE-AngMom}
	\mathbf{P} = \mathring\rho \g^\sharpo \frac{\partial \Psi}{\partial \F} \,,\qquad 
	(\F^{-1}\mathbf{P})^\star=\F^{-1}\mathbf{P}\,.
\end{equation}
Now, the rate of dissipation \eqref{Dissipation1b} is simplified to read
\begin{equation} \label{Dissipation2}
	\dot\eta =  \llangle \operatorname{\mathring{D}iv}\mathbf{P}+ \rho(\mathbf{B} 
	- \mathbf{A}),\mathbf{V} \rrangle_{\g}
	-\dot{\rho}_0 \left[\Psi+\frac{1}{2} \Vert\mathbf V\Vert^2_{\g}\right]
	-\frac{1}{\Theta}\langle d\Theta,\mathbf{Q}\rangle
	\geq 0\,.
\end{equation}
One can choose the velocity vector arbitrarily while its norm $\Vert\mathbf V\Vert_{\g}$ is fixed. This implies that the inequality \eqref{Dissipation2} can hold only if
\begin{equation} 
	\operatorname{\mathring{D}iv}\mathbf{P}+ \mathring\rho\mathbf{B} =\mathring\rho\mathbf{A} \,.
\end{equation}
Now the rate of dissipation takes the following form
\begin{equation} \label{Dissipation3}
	\dot\eta =  -\dot{\rho}_0 \left[\Psi+\frac{1}{2} \Vert\mathbf V\Vert^2_{\g}\right]
	-\frac{1}{\Theta}\langle d\Theta,\mathbf{Q}\rangle
	\geq 0\,.
\end{equation}
Next, one can choose the velocity vector norm $\Vert\mathbf V\Vert_{\g}$ arbitrarily while keeping the other fields fixed. For all these extended thermodynamics processes the above inequality must hold.
This implies that $\dot{\rho}_0=0$ and ${\dot\eta = -\frac{1}{\Theta}\langle d\Theta,\mathbf{Q}\rangle\geq 0}\,$.
\end{proof}

\begin{remark}
\citet{coleman1963} showed that requiring \eqref{Second-Law-Local} to hold for all admissible thermodynamic processes gives the Doyle-Ericksen formula \eqref{Doyle-Ericksen}.
We have shown that requiring \eqref{Second-Law-Local} to hold for all admissible extended thermodynamic processes gives all the balance laws and the Doyle-Ericksen formula \eqref{Doyle-Ericksen}.
It is known that in hyperelaticity, balance of angular momentum implies objectivity \citep{Kadic1980,Yavari2024Cauchy}. Therefore, we have shown that the first and second laws of thermodynamics imply objectivity as well.
\end{remark}

\begin{remark}
For an incompressible hyperelastic solid, the Legendre transform \eqref{Free-Energy} is modified to take into account  the constraint of volume preservation $J=1$ on motions as follows
\begin{equation}\label{Free-Energy_trans_inc}
\Psi -p(J-1) = \mathscr E - \Theta \mathscr N\,,
\end{equation}
where $p(X,t)$ is the Lagrange multiplier associated with the incompressibility constraint.
Computing $\dot J\,$, we find\footnote{We use the identity $\frac{d}{d t}\left[\det\F \right]=\left[\mathbf F^{-\star} \!:\! \bar{\operatorname{D}}_t\F\right]\det\mathbf F\,$, perform a computation similar to \eqref{eq:rate-decomp}, and observe that $\F^{-1} \g^\sharpo \F^{-\star}$ is a symmetric tensor.}
\begin{equation}\label{Free-Energy_trans_inc}
\dot J
= J \mathbf F^{-\star} \!:\! \bar{\operatorname{D}}_t \F
=\left(\F^{-1} \g^\sharpo \F^{-\star}\right) \!:\! (\Ds+ \Da)
=\left(\F^{-1} \g^\sharpo \F^{-\star}\right) \!:\! \Ds\,.
\end{equation}
Revisiting the proof for Proposition~\ref{prop:hyper}, the results remain unchanged except for the Doyle-Ericksen formula \eqref{Doyle-Ericksen} which is modified to read
\begin{equation}
	\mathbf P =  \mathring\rho \g^\sharpo \frac{\partial \Psi}{\partial \F} - p\, \g^\sharpo \F^{-\star}\,.
\end{equation}
\end{remark}

\begin{remark}
Note that it is possible to recast the balance and constitutive equations \eqref{eq:prop_hyper} in their spatial (current) form as follows:
\begin{subequations}\label{eq:spat-prop}
\begin{empheq}[left={\empheqlbrace}]{align}
\label{eq:spat-prop-mass}
& \dot{\rho} + \rho \operatorname{div}\mathbf{v}=0 \,,\\
\label{eq:Cauchy-const}
& \boldsymbol\sigma=\rho\g^\sharpo \frac{\partial \Psi}{\partial \F}\F^\star= 2 \rho \frac{\partial \Psi}{\partial \g} \,,\\
\label{eq:spat-prop-linear-momentum}
& \operatorname{div}\boldsymbol\sigma+ \rho\,\mathbf{b} =\rho\,\mathbf{a} \,,\\
\label{eq:spat-prop-angular-momentum}
& \boldsymbol\sigma^{\star}=\boldsymbol\sigma \,,\\
& \dot\eta = - \frac{1}{\Theta} \langle d\Theta, \mathbf Q \rangle \geq 0 \,,
\end{empheq}
\end{subequations}
where $\rho$ is the spatial mass density, $\boldsymbol{\sigma}$ is the Cauchy stress tensor, $\mathbf{b}=\mathbf{B}\circ\varphi_t^{-1}\,$, and $\operatorname{div}$ denotes the spatial divergence operator, i.e. in the manifold $\left(\mathcal S,\g\right)\,$.
The spatial balance of mass \eqref{eq:spat-prop-mass} is derived from the material (reference) balance and the relation $\rho = \mathring\rho/\mathring{J}\,$, which follows from the definition of the Jacobian, $\varphi^*dv = \mathring{J} d\mathring{V}\,$.
In \eqref{eq:Cauchy-const}, the Cauchy stress tensor $\boldsymbol\sigma$ is the Piola transform of the first Piola-Kirchhoff stress tensor $\mathbf{P}$; consistently with the spatial forms of the balance of linear momentum \eqref{eq:spat-prop-linear-momentum} and angular momentum \eqref{eq:spat-prop-angular-momentum} obtained by pushing forward the respective material equations into the current configuration using the Piola transformation \citep{MarsdenHughes1983}.
\end{remark}
\begin{remark}[Connection with Rational Extended Thermodynamics]
\label{remark:RET2}
As briefly noted earlier in Remark~\ref{remark:RET1}, the term ``extended'' appears prominently in Rational Extended Thermodynamics (RET), developed by M\"uller and Ruggeri \citep{Muller1972,MullerRuggeri1998RET,Ruggeri2012a,Ruggeri2012b,Ruggeri2024}, as well as in our proposed framework, though with distinct meanings and conceptual motivations.
In RET, extension refers to an enlargement of the thermodynamic state space: beyond the classical variables (mass, momentum, energy), RET includes higher-order moments of these quantities as independent fields. This leads to a hyperbolic formulation of continuum thermodynamics, thereby resolving the longstanding paradox of infinite propagation speeds inherent in classical theories \citep{muller1966ausbreitungsgeschwindigkeit, muller1967paradoxon}, and ensuring consistency with kinetic theory \citep{grad1949kinetic,grad1958principles}.
By contrast, our use of ``extended'' pertains to the class of thermodynamic processes considered. Specifically, we define ``extended processes'' as those that satisfy the first law of thermodynamics, without assuming a priori satisfaction of the balance of mass, linear momentum, or angular momentum. In the classical Coleman–Noll framework, such balance laws are assumed from the outset. An ``extended process'' is deemed admissible if it also satisfies the Clausius–Duhem inequality and its constitutive assumptions hold everywhere in the body and at all times. 
We demonstrate that this condition yields both the constitutive relations (Doyle-Ericksen formula) and the full set of balance laws, thereby generalizing the classical Coleman–Noll procedure.
Moreover, our formulation eliminates the need to impose objectivity (observer invariance or material frame indifference) as an axiom. Instead, we find that objectivity arises as a consequence of the first and second laws of thermodynamics—a feature shared with RET.
While RET uses the entropy inequality to constrain the dynamics of an enlarged state space, our approach employs it to derive balance laws and constitutive structures within the conventional setting of continuum mechanics. Despite the differing conceptual starting points, the structural parallels between RET and our framework are notable, and we believe further exploration of these connections may prove fruitful.
\end{remark}

\section{Thermodynamics and the Balance Laws of Hyper-Anelasticity}
\label{Sec:ThermoAnElast}

In this section, we extend the generalized Coleman-Noll procedure to the context of hyper-anelasticity\textemdash a concept we define and elaborate on below.
Anelasticity extends the concept of elasticity by accounting not only for reversible finite deformations under external mechanical forces, as observed in elastic materials, but also for additional, often irreversible, distortions. These distortions---referred to as eigenstrains---originate from internal microstructural reconfigurations driven by physical, chemical, or biological processes. Processes such as temperature changes, growth, remodeling, and defect dynamics can induce such eigenstrains.

As discussed in \S\ref{Sec:NonEuclidG}, the Bilby-Köner-Lee (BKL) decomposition, $\F = \Fe \Fa\,$, provides a framework for separating the recoverable elastic deformation, $\Fe\,$, from the intrinsic anelastic distortion, $\Fa\,$. As a consequence of this decomposition, the material metric $\G = \Fa^* \Go$ emerges to describe the evolving, non-Euclidean geometry of the  stress-free material configuration, reflecting the intrinsic distortions arising from eigenstrains.
While elastic constitutive modelling requires only a stress-strain relationship, anelastic constitutive modelling goes further by accounting for microstructural reconfigurations. Anelasticity thus necessitates additional constitutive assumptions to describe these internal rearrangements. In the context of the BKL decomposition discussed above, an extra constitutive device is needed to prescribe the evolution of the anelastic distortion $\Fa$, typically introduced through a configurational force, which acts as an internal driving force for microstructural dynamics \citep{Maugin2010ConfForces}. 

\begin{defi}
Within this framework, we define a \textit{hyper-anelastic} material as one in which the stress derives from a scalar strain energy function.\footnote{Note, however, that for a hyper-anelastic material, the configurational forces are not necessarily derived from a scalar potential.}We further identify a subclass of hyper-anelastic materials, which we term \textit{Rayleigh hyper-anelastic}, characterized by the existence of a Rayleigh dissipation potential that governs the evolution of the configurational forces.
It should however be noted that not all anelastic materials are hyper-anelastic. 
\end{defi}

Similarly to the discussion on elasticity in Remark~\ref{rmrk:hyper}, a larger class of anelastic materials is one in which the stress at a material point depends only on the elastic distortion at that point, without any history dependence, i.e. $\boldsymbol{\sigma} = \boldsymbol{\mathsf{f}}(\be)\,$---\textit{Cauchy anelasticity}, where $\be$ is the elastic Finger tensor. 
An even larger class is \textit{implicit anelasticity} where the stress-strain relationship is defined implicitly, i.e. $\boldsymbol{\mathsf{f}}(\boldsymbol{\sigma}, \be) = \mathbf{0}\,$. 

Building on this foundation, we use the first and second laws of thermodynamics for anelasticity, incorporating the interplay between elastic and anelastic distortions, and make use of the extended thermodynamic process concept previously introduced in Definition~\ref{def:Extended}.
This generalized framework enables the derivation of constitutive equations, balance laws, and thermodynamic constraints governing hyper-anelastic materials, all without invoking any assumption of invariance.

\subsection{The First Law of Thermodynamics}

For an anelastic solid, we write energy balance as \citep{EpsteinMaugin2000, LubardaHoger2002}\footnote{Note that the last term in \eqref{eq:An-Thermo_First} is added to account for the change in the internal and kinetic energies of the system due to bulk growth or resorption with a material rate of change of mass $S_m\,$.}

\begin{equation}\label{eq:An-Thermo_First}
\begin{split}
	\frac{d}{ dt}\int_{\mathcal{U}} \rho\left(\mathscr{E}+\frac{1}{2} \Vert\mathbf V\Vert^2_{\mathbf g}\right)
	 dV
	=\int_{\mathcal{U}}\rho \left(\llangle\mathbf{B},\mathbf{V}\rrangle_{\mathbf{g}}
	+R\right) dV+\int_{\partial \mathcal{U}}
	\left(\llangle\mathbf{T},\mathbf{V}\rrangle_{\mathbf{g}}-\llangle \mathbf{Q},\mathbf{N}\rrangle_{\mathbf{G}}\right)dA\\
	+\int_{\mathcal{U}} S_m\left(\mathscr{E}+\frac{1}{2} \Vert\mathbf V\Vert^2_{\mathbf g}\right)
	 dV\,,
\end{split}
\end{equation}
for any open set $\mathcal{U}\subset\mathcal{B}\,$, where $\mathscr{E}$ is the specific internal energy, $\rho$ is the material mass density, ${S_m=S_m(X,t)}$ is the material rate of change of mass per unit (stress-free) volume\textemdash it is identically equal to zero in the absence of bulk growth or resorption, $\mathbf{B}$ is the specific body force, $\mathbf{T} = \mathbf{T} (X, \mathbf N)$ is the boundary traction vector field per unit (stress-free) material area, $\mathbf{N}$ is the $\mathbf{G}$-unit normal to the boundary $\partial\mathcal U\,$, ${R=R(X,t)}$ is the external specific heat supply, ${\mathbf Q=\mathbf Q(X,\Theta,d\Theta,\mathbf C,\mathbf G)}$ is the external heat flux per unit material (stress-free) area and, $dA$ is the material area element.

By the same argument developed in \S\ref{Sec:1stLaw-Elast}, using Marsden and Hughes's version of Cauchy's theorem \citep[\S2.1(1.9)]{MarsdenHughes1983}, we find $\mathbf{T} = \mathbf{P} \mathbf{N}^\flat\,$, where $\mathbf{P}$ is a second-order two-point tensor, and $(.)^\flat$ denotes the musical isomorphisms for lowering with respect to $\G$ and $\g\,$. Hence, the energy balance \eqref{eq:Thermo_First} in local form reads:\footnote{
Recall that $\mathbf{G}=\Fa^*\mathring{\mathbf{G}}$ and $ dV=\sqrt{\det\mathbf{G}}\, dX^1\wedge dX^2\wedge dX^3\,$. Hence, we ought to consider the implicit time dependance of $ dV$ through $\det\mathbf{G}$ in the left-hand-side of \eqref{eq:An-Thermo_First} and we use the identity $\frac{d}{ d t}\left[\det\mathbf G \right]=\left[\dot{\mathbf G}\!:\!\mathbf G^{-1}\right]\det\mathbf G\,$.}
\begin{equation}\label{eq:An-loc_Thermo_First}
\begin{split}
	\rho\,\dot{\mathscr{E}} &= (\mathbf F^{-1} \mathbf P) \!:\! \left(\Ds +\Da\right) +\rho R 
	- \operatorname{Div} \mathbf Q
	+ \llangle \operatorname{Div}\mathbf{P}+ \rho(\mathbf{B} 
	- \mathbf{A}),\mathbf{V} \rrangle_{\mathbf{g}}
	\\&\quad
	+ \left(S_m-\dot{\rho} - \frac{1}{2}\rho\,\dot{\mathbf G}\!:\!\mathbf G^{\sharp} \right)\left(\mathscr{E}+\frac{1}{2} \Vert\mathbf V\Vert^2_{\mathbf g}\right)\,,
\end{split}
\end{equation}
where, in the term $\operatorname{Div} \mathbf{Q}$, $\operatorname{Div}$ denotes the material divergence with respect to the connection $\nabla$, while in the term $\operatorname{Div} \mathbf{P}$, it denotes the two-point divergence operator with respect to the connections $\nabla$ and $\bar{\nabla}$ (more precisely, the divergence with respect to the induced connection), $(.)^\sharp$ denotes the musical isomorphisms for raising indices with respect to $\G$ and $\g\,$, and $\Ds$ and $\Da$ are the symmetric and anti-symmetric material rate of deformation tensors, respectively, as defined in \eqref{Eq:Def_rates}.

\subsection{The Second Law of Thermodynamics}

For an anelastic solid, we write the Clausius-Duhem inequality in the anelastic material manifold as \citep{EpsteinMaugin2000,LubardaHoger2002}\footnote{Note that the last term in \eqref{eq:Thermo_Second} is added to account for the change in the entropy of the system due to bulk growth or resorption with a material rate of change of mass $S_m\,$.}
\begin{equation}\label{eq:An-Thermo_Second}
	\frac{d}{ dt} \int_{\mathcal{U}} \rho \mathscr{N} dV
	\geq\int_{\mathcal{U}}\rho \frac{R}{\Theta} dV
	-\int_{\partial\mathcal{U}}\frac{1}{\Theta}\llangle \mathbf{Q},\mathbf{N}\rrangle_{\G}dA
	+\int_{\mathcal{U}} S_m\, \mathscr{N} dV\,,
\end{equation}
for any open set $\mathcal{U}\subset\mathcal{B}\,$, where $\mathscr{N}$ is the specific entropy, i.e. entropy per unit mass. In localized form, the material Clausius-Duhem inequality \eqref{eq:An-Thermo_Second} is written as
\begin{equation} \label{An-loc_Thermo_Second}
	\dot\eta = \rho\, \dot{\mathscr{N}}\Theta + \Theta\operatorname{Div}\left(\frac{\mathbf Q}{\Theta}\right) - \rho R
	- \left(S_m-\dot{\rho} - \frac{1}{2}\rho\,\dot{\mathbf G}\!:\!\mathbf G^{\sharp} \right) \Theta \mathscr N \geq 0\,.
\end{equation}

\subsection{Constitutive Equations and Balance Laws in Hyper-Anelasticity}

As discussed in \S\ref{Sec:NonEuclidG}, intrinsic distortions (eigenstrains) give rise to a non-Euclidean stress-free configuration, represented by the incompatible distortion $\Fa\,$. The recoverable elastic distortion $\Fe$ then maps this stress-free configuration to the current spatial configuration. Consequently, the free energy for a hyper-anelastic material depends on $\Fe$ rather than the full deformation gradient $\F\,$, i.e. $\Psi = {\Psi}(X,\Theta,\Fe,\Go,\g)\,$. As the Legendre transform of the specific internal energy $\mathscr E$ with respect to the conjugate variables temperature $\Theta$ and specific entropy $\mathscr N\,$, the specific free energy function reads
\begin{equation}\label{An-Free-Energy}
	\Psi = \mathscr E - \Theta \mathscr N\,.
\end{equation}
Hence, we find that ${\mathscr E = {\mathscr E}(X, \mathscr N,\Fe,\Go,\g)}$ and 
\begin{equation}
	\mathscr N = -\frac{\partial \Psi}{\partial \Theta}\,.
\end{equation}

\begin{remark}
The material symmetry group $\mathring{\mathcal G}_X$ of a hyper-anelastic material (constitutively defined by a free energy $\Psi$) at a point $X\in\mathcal{B}$ with respect to the Euclidean reference configuration $(\mathcal{B},\mathring{\mathbf{G}})$ is the set of $\mathring{\mathbf{K}}\in\mathrm{Orth}(\mathring{\mathbf{G}})=\{\mathbf{Q}: T_X\mathcal{B}\to T_X\mathcal{B}~|~ {\mathbf{Q}}^{\star}\mathring{\mathbf{G}}{\mathbf{Q}}=\mathring{\mathbf{G}} \}$ such that
\begin{equation} \label{Elasticity-Sym-Group}
\mathring{\mathbf{K}}_*\Psi(X,\Theta,\Fe,\mathring{\mathbf{G}},\mathbf{g})=
	\Psi(X,\Theta,\mathring{\mathbf{K}}^*\Fe,\mathring{\mathbf{K}}^*\mathring{\mathbf{G}},\mathbf{g})
	=\Psi(X,\Theta,\Fe,\mathring{\mathbf{G}},\mathbf{g})
	\,,
\end{equation}
for any elastic distortion $\Fe\,$.
Introducing a set of structural tensors $\mathring{\boldsymbol\Lambda}$ for $\mathring{\mathcal G}_X$\footnote{The subgroup $\mathring{\mathcal G}_X \leqslant\mathrm{Orth}(\mathring{\mathbf{G}})$ can be characterized by a finite collection of \emph{structural tensors} $\mathring{\boldsymbol{\Lambda}}_i\,$, $i=1,\dots,N\,$, which is a basis for the space of $\mathring{\mathcal G}_X$-invariant tensors~\citep{liu1982, boehler1987, zheng1993, zheng1994theory, lu2000covariant, MazzucatoRachele2006}.} as additional independent variables of the free energy, it follows that $\Psi=\Psi(X,\Theta,\Fe,\mathring{\boldsymbol\Lambda},\mathring{\mathbf{G}},\mathbf{g})$ is materially covariant, i.e. for any linear transformation $\boldsymbol{\mathsf{T}}:T_X\mathcal{B}\to T_X\mathcal{B}\,$, ${\Psi}(X,\Theta,\Fe,\mathring{\boldsymbol\Lambda},\mathring{\mathbf{G}},\mathbf{g})={\Psi}(X,\Theta,\boldsymbol{\mathsf{T}}^*\Fe,\boldsymbol{\mathsf{T}}^*\mathring{\boldsymbol\Lambda},\boldsymbol{\mathsf{T}}^*\mathring{\mathbf{G}},\mathbf{g})\,$. If we take $\boldsymbol{\mathsf{T}}=\Fa\,$, and recalling that $\Fa^*\Fe=\Fe\Fa=\mathbf{F}$ and $\G=\Fa^*\Go=\Fa^\star\Go\Fa\,$, it follows that
\begin{equation} \label{Eq:FreeAnelast}
	\Psi
	={\Psi}(X,\Theta,\Fe,\mathring{\boldsymbol\Lambda},\mathring{\mathbf{G}},\mathbf{g})
	={\Psi}(X,\Theta,\mathbf{F},\boldsymbol\Lambda,\mathbf{G},\mathbf{g})
	\,,
\end{equation}
where $\boldsymbol\Lambda=\Fa^*\mathring{\boldsymbol\Lambda}$ is the set of structural tensors in $(\mathcal B,\G)\,$. We see in \eqref{Eq:FreeAnelast} how the material metric $\G=\Fa^*\Go$ naturally emerges following the hyper-anelastic constitutive assumption.
\end{remark}
 
\begin{prop} \label{prop:hyperAn}
For a hyper-anelastic body, the first and second laws of thermodynamics \eqref{eq:An-loc_Thermo_First} and \eqref{An-loc_Thermo_Second} imply that
\begin{subequations}\label{eq:prop_hyperAn}
\begin{empheq}[left={\empheqlbrace}]{align}
	\label{An-Mass-Conservation}
	& \dot{\rho} + \frac{1}{2}\rho\,\dot{\mathbf G}\!:\!\mathbf G^{-1} = S_m \,, \\
	\label{An-Doyle-Ericksen}
	& \mathbf P=\rho\g^\sharp \frac{\partial \Psi}{\partial\Fe}\Fa^{-\star}\,,\\
	\label{An-Linear-Momentum}
	& \operatorname{Div}\mathbf{P}+ \rho \mathbf{B} = \rho\mathbf{A} \,, \\
	\label{An-Angular-Momentum}
	& \F\mathbf{P}^{\star}=\mathbf{P}\F^{^\star}\,,\\
	\label{An-Diss_Ineq}
	& \dot\eta =
	\rho \Fe^\star\frac{\partial \Psi}{\partial \Fe} \!:\! \dot\Fn \Fn^{-1}
	+ \rho \Fn^\star \Fe^\star\frac{\partial \Psi}{\partial \Fe}  \Fn^{-\star} \!:\! \boldsymbol\alpha \dot\Theta
	- \frac{1}{\Theta} \langle d\Theta, \mathbf Q \rangle
	\geq 0\,.
\end{empheq}
\end{subequations}
In other words, the first and second laws of thermodynamics imply the balance of mass \eqref{An-Mass-Conservation}, the Doyle-Ericksen formula \eqref{An-Doyle-Ericksen}, the balance of linear momentum \eqref{An-Linear-Momentum}, the balance of angular momentum \eqref{An-Angular-Momentum}, and the dissipation inequality \eqref{An-Diss_Ineq}. Note that \eqref{An-Doyle-Ericksen} effectively shows that the two-point tensor $\mathbf{P}$ is indeed the first Piola-Kirchhoff stress tensor.
\end{prop}
\begin{proof}
Using \eqref{eq:An-loc_Thermo_First}, \eqref{An-loc_Thermo_Second}, and \eqref{An-Free-Energy}, one finds
\begin{equation}\label{eq:An-Dissipation1}
\begin{split}
	\dot\eta =
	(\mathbf F^{-1} \mathbf P) \!:\! \left(\Ds +\Da\right)
	-\rho\dot{\Psi}
	-\rho \dot{\Theta} \mathscr{N}
	+ \llangle \operatorname{Div}\mathbf{P}+ \rho(\mathbf{B} 
	- \mathbf{A}),\mathbf{V} \rrangle_{\mathbf{g}}
	&\\
	+ \left(S_m-\dot{\rho} - \frac{1}{2}\rho\,\dot{\mathbf G}\!:\!\mathbf G^{\sharp} \right)\left(\Psi+\frac{1}{2} \Vert\mathbf V\Vert^2_{\mathbf g}\right)
	- \frac{1}{\Theta} \langle d\Theta, \mathbf Q \rangle
	&\geq 0\,.
\end{split}
\end{equation}
Expanding $\dot\Psi$ by Leibniz rule and using the product rule $\bar{\operatorname{D}}_t \Fa = (\bar{\operatorname{D}}_t \F) \Fa^{-1} - \Fe \dot\Fa \Fa^{-1}\,$,\footnote{Unlike the two-point tensors $\F$ and $\Fe\,$, which are sections of the product of the tangent bundle of the material manifold $\mathcal{B}$ and the cotangent bundle of the spatial manifold $\varphi_t(\mathcal{B})$, the referential tensor $\Fa$ is a section of the product of the tangent and cotangent bundles of the fixed material manifold $\mathcal{B}\,$. As a result, the total time derivatives of $\F$ and $\Fe$ are not well-defined due to the time-dependent nature of the spatial manifold $\varphi_t(\mathcal{B})$, and their material rates must instead be expressed using the spatial covariant derivative $\bar{\operatorname{D}}_t(.)$ to account for the evolving geometry of the spatial configuration. In contrast, the material rate of $\Fa$ is given by its well-defined total time derivative, $\dot{\Fa} \coloneq {d\Fa}/{dt}\,$, since it is defined entirely on the fixed material manifold $\mathcal{B}\,$.}
 one obtains
\begin{equation}\label{eq:An-psi-dot}
\begin{split}
	\dot{\Psi}
	&=\frac{\partial \Psi}{\partial \Theta}\dot{\Theta}
	+\frac{\partial \Psi}{\partial \Fe}\!:\!\bar{\operatorname{D}}_t \Fe
	+\frac{\partial \Psi}{\partial \g}\!:\!\bar{\operatorname{D}}_t \g\\
	&=\frac{\partial \Psi}{\partial \Theta}\dot{\Theta}
	+\frac{\partial \Psi}{\partial \Fe}\!:\!\left[
	(\bar{\operatorname{D}}_t \F) \Fa^{-1} - \Fe \dot\Fa \Fa^{-1}
	\right]\\
	&=\frac{\partial \Psi}{\partial \Theta}\dot{\Theta}
	+\frac{\partial \Psi}{\partial \Fe} \Fa^{-\star}\!:\!\bar{\operatorname{D}}_t\F
	-\Fe^\star\frac{\partial \Psi}{\partial \Fe}\!:\! \dot\Fa \Fa^{-1}\\
	&=-\mathscr N \dot{\Theta}
	+\F^{-1} \g^\sharp\frac{\partial \Psi}{\partial \Fe} \Fa^{-\star}\!:\!(\Da +\Ds)
	-\Fe^\star\frac{\partial \Psi}{\partial \Fe}\!:\! \dot\Fa \Fa^{-1}
	\,,
\end{split}
\end{equation}
where we use metric compatibility to write $\bar{\operatorname{D}}_t \g = \mathbf 0\,$, recall that $\frac{\partial \Psi}{\partial \Theta}=-\mathscr N\,$, and similarly to \eqref{eq:rate-decomp}, we use the following identity:
\begin{equation}
\frac{\partial \Psi}{\partial \Fe} \Fa^{-\star}\!:\!\bar{\operatorname{D}}_t\F = \F^{-1} \g^\sharp\frac{\partial \Psi}{\partial \Fe} \Fa^{-\star}\!:\!(\Da +\Ds)\,.
\end{equation}
Similarly to \eqref{eq:F-symmetry}, it can be shown that $\F^{-1} \g^\sharp({\partial \Psi}/{\partial \Fe}) \Fa^{-\star}$ is a symmetric tensor, which then implies that
\begin{equation}
\F^{-1} \g^\sharp\frac{\partial \Psi}{\partial \Fe} \Fa^{-\star}\!:\!\Da=0\,.\end{equation}
Now substituting \eqref{eq:An-psi-dot} into \eqref{eq:An-Dissipation1}, one finds
\begin{equation}\label{eq:An-Dissipation2}
\begin{split}
	\dot\eta =
	\left[\mathbf F^{-1}\left( \mathbf P- \rho\g^\sharp\frac{\partial \Psi}{\partial \Fe} \Fa^{-\star}\right)\right] \!:\! \Ds
	+(\mathbf F^{-1} \mathbf P) \!:\! \Da
	+\rho \Fe^\star\frac{\partial \Psi}{\partial \Fe} \!:\! \dot\Fa \Fa^{-1}
	- \frac{1}{\Theta} \langle d\Theta, \mathbf Q \rangle
	&\\
	+ \llangle \operatorname{Div}\mathbf{P}+ \rho(\mathbf{B} 
	- \mathbf{A}),\mathbf{V} \rrangle_{\mathbf{g}}
	+ \left(S_m-\dot{\rho} - \frac{1}{2}\rho\,\dot{\mathbf G}\!:\!\mathbf G^{\sharp} \right)\left(\Psi+\frac{1}{2} \Vert\mathbf V\Vert^2_{\mathbf g}\right)
	&\geq 0\,.
\end{split}
\end{equation}
This inequality must hold for all motions, i.e. all extended thermodynamic processes. Similarly to the proof of Proposition~\ref{prop:hyper}, it follows from \eqref{eq:An-Dissipation2}, by arbitrariness and independence of $\Ds\,$, $\Da\,$, $\mathbf V/\Vert\mathbf V\Vert^2_{\mathbf g}\,$, and $\Vert\mathbf V\Vert^2_{\mathbf g}\,$, that
\begin{subequations}
\begin{align}
&	\mathbf P = \rho\g^\sharp\frac{\partial \Psi}{\partial \Fe} \Fa^{-\star}\,,\\
&	(\mathbf F^{-1} \mathbf P)^\star = \mathbf F^{-1} \mathbf P\,,\\
&	\operatorname{Div}\mathbf{P}+ \rho\mathbf{B} = \rho \mathbf{A}\,,\\
&	\dot{\rho} + \frac{1}{2}\rho\,\dot{\mathbf G}\!:\!\mathbf G^{\sharp}  = S_m\,,
\end{align}
\end{subequations}
respectively. Consequently, \eqref{eq:An-Dissipation2} simplifies to
\begin{equation}
\label{eq:An-dissip}
	\dot\eta =
	\rho \Fe^\star\frac{\partial \Psi}{\partial \Fe} \!:\! \dot\Fa \Fa^{-1}
	- \frac{1}{\Theta} \langle d\Theta, \mathbf Q \rangle
	\geq 0\,.
\end{equation}
Recalling the decomposition $\Fa=\Fn\Fth\,$, we may rewrite the first term in \eqref{eq:An-dissip} as
\begin{equation}\label{eq:decoAnDiss}
\begin{split}
\Fe^\star\frac{\partial \Psi}{\partial \Fe} \!:\! \dot\Fa \Fa^{-1}
	&= \Fe^\star\frac{\partial \Psi}{\partial \Fe} \!:\! \dot\Fn \Fn^{-1}
+ \Fe^\star\frac{\partial \Psi}{\partial \Fe} \!:\! \Fn \dot\Fth \Fth^{-1} \Fn^{-1}\\
	&= \Fe^\star\frac{\partial \Psi}{\partial \Fe} \!:\! \dot\Fn \Fn^{-1}
+ \Fn^\star \Fe^\star\frac{\partial \Psi}{\partial \Fe}  \Fn^{-\star} \!:\! \dot\Fth \Fth^{-1}\\
	&= \Fe^\star\frac{\partial \Psi}{\partial \Fe} \!:\! \dot\Fn \Fn^{-1}
+ \Fn^\star \Fe^\star\frac{\partial \Psi}{\partial \Fe}  \Fn^{-\star} \!:\! \boldsymbol\alpha \dot\Theta\,,
\end{split}
\end{equation}
where following \eqref{eq:Th_distort} we use $ \dot\Fth\Fth^{-1}=\boldsymbol\alpha \dot\Theta\,$. This then allows us to isolate the anelastic rate of dissipation in \eqref{eq:An-dissip} into an athermal anelastic contribution and a purely thermal contribution as 
\begin{equation}\label{An-Diss_Ineq_proof}
	\dot\eta =
	\rho \Fe^\star\frac{\partial \Psi}{\partial \Fe} \!:\! \dot\Fn \Fn^{-1}
	+ \rho \Fn^\star \Fe^\star\frac{\partial \Psi}{\partial \Fe}  \Fn^{-\star} \!:\! \boldsymbol\alpha \dot\Theta
	- \frac{1}{\Theta} \langle d\Theta, \mathbf Q \rangle
	\geq 0\,.
\end{equation}
\end{proof}

\begin{remark}
In formulating the laws of thermodynamics for anelsticity, previous works, such as \citep{EpsteinMaugin2000, LubardaHoger2002, Yavari2010, sadik2015geometric}, introduced additional terms accounting for the time dependence of the material geometry\footnote{Recall that since the material metric is given by $\mathbf{G} = \Fa^* \Go\,$, it is indeed implicitly time dependant via the time dependance of $\Fa\,$.} on the right-hand side of the energy balance \eqref{eq:An-Thermo_First} and the Clausius-Duhem inequality \eqref{eq:An-Thermo_Second}.\footnote{\citet{EpsteinMaugin2000} introduce $\Pi_\varepsilon$ and $M_\varepsilon$ in Eq. (7.1) \& (8.1) as ``the ``irre-versible'' volume and surface contributions to the internal energy,'' while \citet{LubardaHoger2002} introduce $\mathscr{R}_g r_g$ in Eq. (5.3) as ``the rate of chemical energy per unit current mass.''
In \citep[Eq. (2.18) \& (2.37)]{Yavari2010} and \citep[Eq. (3.26) \& (3.30))]{sadik2015geometric}, these terms appear in the form $({\partial \mathscr{E}}/{\partial \mathbf{G}})\!:\!\dot\G$ and $({\partial \mathscr{N}}/{\partial \mathbf{G}})\!:\!\dot\G\,$, respectively, to account for the time dependance of the material metric.}
It turns out however that adding such a ``correction term'' would force the anelastic process to be non-dissipative. Indeed, the first term in the dissipation inequality \eqref{eq:An-dissip}, which account precisely for the anelastic dissipation, would effectively cancel against the a priori correction if it were to be included, resulting in excluding anelastic dissipation from the model.
Therefore, our proposed formulation herein correctly captures the energy dissipation intrinsic to evolving material geometries without introducing any additional artificial compensations.
\end{remark}

\begin{remark}
For an incompressible hyper-anelastic solid, the Legendre transform \eqref{Free-Energy} is modified to take into account the constraints of elastic incompressibility ($\Je=1$) and/or athermal anelastic incompressibility ($\Jn=1$) as\footnote{Thermal effects can indeed induce volumetric changes; however, the evolution of the temperature field is fully determined by the heat equation. Imposing an additional constraint, such as “thermal incompressibility,” on the temperature field is therefore not physically justified.}
\begin{equation}\label{Free-Energy_trans_inc}
\rho\Psi -\pe(\Je-1) - \pn(\Jn-1)= \rho \mathscr E - \Theta \rho \mathscr N\,,
\end{equation}
where $\pe(X,t)$ and $\pn(X,t)$ are the Lagrange multipliers associated with the respective incompressibility constraints.
Computing $\dot\Je$ and $\dot\Jn\,$, we find
\begin{equation}\label{eq:Jac_diff}
	\dot \Je =
	\Je \left(\F^{-1} \g^\sharp \F^{-\star}\right) \!:\! \Ds
	- \Je\Fn^{-\star}\!:\!\dot\Fn - \Je (\mathbf{I}\!:\!\boldsymbol\alpha) \dot{\Theta}\,,\qquad
	\dot\Jn= \Jn \,\Fn^{-\star}\!:\!\dot\Fn\,.
\end{equation}
Revisiting the proof for Proposition~\ref{prop:hyperAn}, the results remain unchanged except for the Doyle-Ericksen formula \eqref{An-Doyle-Ericksen} which is modified to read
\begin{equation}
	\mathbf P =  \rho \g^\sharp \frac{\partial \Psi}{\partial \Fe} \Fa^{-\star} - \pe\, \g^\sharp \F^{-\star}\,,
\end{equation}
and the dissipation inequality \eqref{An-Diss_Ineq}, which is modified to read
\begin{equation}
\dot\eta =
	\left(\rho \Fe^\star\frac{\partial \Psi}{\partial \Fe} - (\pe-\pn)\mathbf I \right) \!:\! \dot\Fn \Fn^{-1}
	+ \left(\rho \Fn^\star \Fe^\star\frac{\partial \Psi}{\partial \Fe}  \Fn^{-\star} -\pe \mathbf I \right)\!:\! \boldsymbol\alpha \dot\Theta
	- \frac{1}{\Theta} \langle d\Theta, \mathbf Q \rangle
	\geq 0\,.
\end{equation}
\end{remark}

\subsection{Thermal and Kinetic Evolution Equations in Rayleigh Hyper-Anelasticity}

\subsubsection{Heat equation}

Let us first derive the heat equation to characterise the evolution of the temperature field in the case of a non-isothermal process. Using the balance laws derived in the previous subsection, and using the Legendre transform, the localised energy balance \eqref{eq:An-loc_Thermo_First} reads 
\begin{equation}\label{eq:HeatEq1}
\rho\,\dot{\Psi}+\rho\,\dot{\Theta}\mathscr{N}+\rho\,\Theta\dot{\mathscr{N}}  = (\mathbf F^{-1} \mathbf P) \!:\! \Ds +\rho R - \operatorname{Div} \mathbf Q\,.
\end{equation}
Recalling \eqref{eq:An-psi-dot} and \eqref{eq:decoAnDiss}, we may write
\begin{subequations}\label{eq:HeatEq2}
\begin{align}
	\dot{\Psi}
	&= - \mathscr{N}\dot{\Theta}
	+ \frac{\partial \Psi}{\partial \Fe} \Fa^{-\star} \!:\! \bar{\operatorname{D}}_t\F
	- \Fe^\star\frac{\partial \Psi}{\partial \Fe} \!:\! \dot\Fn \Fn^{-1}
	- \Fn^\star \Fe^\star\frac{\partial \Psi}{\partial \Fe}  \Fn^{-\star} \!:\! \boldsymbol\alpha \dot\Theta\,,\\
	\begin{split}
	\dot{\mathscr{N}}
	&= - \frac{d}{dt}{\frac{\partial\hat{\Psi}}{\partial \Theta}}
	= - \frac{\partial^2 \Psi}{\partial \Theta^2}\dot{\Theta}
	- \frac{\partial^2 \Psi}{\partial \Fe \partial \Theta} \Fa^{-\star} \!:\! \bar{\operatorname{D}}_t\F
	+ \Fe^\star\frac{\partial^2 \Psi}{\partial \Fe \partial \Theta} \!:\! \dot\Fn \Fn^{-1}
	\\&\quad
	+ \Fn^\star \Fe^\star\frac{\partial^2 \Psi}{\partial \Fe \partial \Theta}  \Fn^{-\star} \!:\! \boldsymbol\alpha \dot\Theta\,.
	\end{split}
\end{align}
\end{subequations}
Substituting \eqref{eq:HeatEq2} into \eqref{eq:HeatEq1} and using the Doyle-Ericksen formula \eqref{An-Doyle-Ericksen}, one finds
\begin{equation}\label{eq:HeatEq3}
\begin{split}
	- \rho \left[
	\Theta \frac{\partial^2 \Psi}{\partial \Theta^2}
	+ \Fn^\star \Fe^\star \left(\frac{\partial \Psi}{\partial \Fe}
	- \Theta\frac{\partial^2 \Psi}{\partial \Fe \partial \Theta} \right) \Fn^{-\star} \!:\! \boldsymbol\alpha
	\right]\dot\Theta 
	- \rho \Fe^\star\left[ \frac{\partial \Psi}{\partial \Fe}
	- \Theta \frac{\partial^2 \Psi}{\partial \Fe \partial \Theta} \right]\!:\! \dot\Fn \Fn^{-1}
	\\
	- \rho\,\Theta \frac{\partial^2 \Psi}{\partial \Fe \partial \Theta} \Fa^{-\star} \!:\! \bar{\operatorname{D}}_t\F
	= \rho R - \operatorname{Div} \mathbf Q\,.
\end{split}
\end{equation}
We introduce the specific heat capacity at constant strain, and denote it by $c_E\,$, as the quantity of heat required to produce a unit temperature increase in a unit mass of material at constant strain. It is given by the following equation
\begin{equation}\label{eq:c_E_def}
\operatorname{Div} \mathbf{Q} = -\rho c_E \dot\Theta\,.
\end{equation}
It follows by identification of \eqref{eq:c_E_def} with \eqref{eq:HeatEq3} at constant strain, i.e. $\bar{\operatorname{D}}_t\F=\dot{\Fn}=\mathbf{0}\,$, and in the absence of external specific heat supply, i.e. $R=0\,$, that
\begin{equation}\label{eq:c_E}
c_E = 
	-\Theta \frac{\partial^2 \Psi}{\partial \Theta^2}
	- \Fn^\star \Fe^\star \left(\frac{\partial \Psi}{\partial \Fe}
	- \Theta\frac{\partial^2 \Psi}{\partial \Fe \partial \Theta} \right) \Fn^{-\star} \!:\! \boldsymbol\alpha\,.
\end{equation}
Therefore, we find the heat equation for a hyper-anelastic solid as follows
\begin{equation}\label{eq:HeatEq}
	\rho c_E \dot\Theta 
	= - \operatorname{Div} \mathbf Q
	+ \rho\,\Theta \frac{\partial^2 \Psi}{\partial \Fe \partial \Theta} \Fa^{-\star} \!:\! \bar{\operatorname{D}}_t\F
	+ \rho \Fe^\star\left[ \frac{\partial \Psi}{\partial \Fe} - \Theta \frac{\partial^2 \Psi}{\partial \Fe \partial \Theta} \right]\!:\! \dot\Fn \Fn^{-1}
	+ \rho R\,.
\end{equation}

\subsubsection{Configurational forces}

In a hyper-anelastic solid, the rate of anelastic energy dissipation may be written as the sum of athermal and thermal contributions \citep{LeTallec1993, Maugin2010ConfForces}
\begin{equation}\label{eq:dissipation}
	\dot\eta = - \Bn\!:\!\dot{\Fn} - \Bth \dot\Theta
	- \frac{1}{\Theta} \langle d\Theta, \mathbf Q \rangle\,,
\end{equation}
where $\Bn$ and $\Bth$ are the generalized configurational forces governing the athermal and thermal anelastic distortions, respectively.
Noting the arbitrariness of the independent variables $\dot{\Fn}$ and  $\dot{\Theta}\,$, it follows by identification of \eqref{An-Diss_Ineq} and \eqref{eq:dissipation} that
\begin{subequations}\label{eq:Config}
\begin{align}
\label{eq:Config_n}
	\Bn & = - \rho \Fe^\star\frac{\partial \Psi}{\partial \Fe} \Fn^{-\star}
	\,,\\
\label{eq:Config_th}
	\Bth & = - \rho \Fn^\star \Fe^\star\frac{\partial \Psi}{\partial \Fe}  \Fn^{-\star} \!:\! \boldsymbol\alpha
	\,.
\end{align}
\end{subequations}
For an incompressible solid, these are modified to read
\begin{subequations}
\begin{align}
\label{eq:Config_n}
	\Bn & = - \rho \Fe^\star\frac{\partial \Psi}{\partial \Fe} \Fn^{-\star} + (\pe-\pn) \Fn^{-\star}
	\,,\\
\label{eq:Config_th}
	\Bth & = - \rho \Fn^\star \Fe^\star\frac{\partial \Psi}{\partial \Fe}  \Fn^{-\star} \!:\! \boldsymbol\alpha +\pe \operatorname{tr}\boldsymbol\alpha
	\,.
\end{align}
\end{subequations}
\begin{remark}
It can be seen from the conservation of mass \eqref{An-Mass-Conservation} that $\rho=\rho(X,\G,t)\,$, which then allows one to write ${\partial \rho}/{\partial \Theta}=0\,$.\footnote{Note, however, that its total derivative does not necessarily vanish, since it implicitly depends on temperature via the material metric $\G\,$. Its total derivative is computed as follows
\begin{equation}
\frac{d\rho}{d\Theta}
=\frac{d\rho}{d\G} \!:\! \frac{d\G}{d\Theta} 
= \frac{d\rho}{d\G} \!:\! \frac{\partial(\Fa^\star\Go\Fa)}{\partial\Theta}
= \frac{d\rho}{d\G} \!:\! \left(
\boldsymbol\alpha^\star \Fa^\star\Go\Fa
+\Fa^\star\Go\Fa \boldsymbol\alpha \right)
= 2 \frac{d\rho}{d\G} \!:\! \G \boldsymbol\alpha
= 2 \frac{d\rho}{d\G} \!:\! \boldsymbol\alpha^\flat\,,
\end{equation}
where recalling \eqref{eq:Th_distort}, we use ${\partial\Fth}/{\partial\Theta} = \Fth \boldsymbol\alpha \,$.} Hence, we may write that
\begin{subequations}
\begin{align}
\begin{split}
\rho \frac{\partial^2 \Psi}{\partial \Fe \partial \Theta} \Fa^{-\star}
& = \frac{ \partial}{ \partial \Theta}\left[\rho \frac{\partial \Psi}{\partial \Fe} \right]\Fa^{-\star}
= \frac{ \partial}{ \partial \Theta}\left[\rho \frac{\partial \Psi}{\partial \Fe} \Fa^{-\star} \right] - \rho \frac{\partial \Psi}{\partial \Fe} \Fn^{-\star} \frac{\partial \Fth^{-\star}}{ \partial \Theta}\\
&= \g \frac{ \partial \mathbf P}{\partial \Theta} - \rho \frac{\partial \Psi}{\partial \Fe} \Fn^{-\star} \Fth^{-\star}\boldsymbol\alpha^{-\star}
= \g \frac{ \partial \mathbf P}{\partial \Theta} - \g \mathbf P \boldsymbol\alpha^{-\star}
\,,
\end{split}\\
\rho\Fe^\star\frac{\partial^2 \Psi}{\partial \Fe \partial \Theta} \Fn^{-\star}
& = \frac{ \partial}{ \partial \Theta}\left[\rho \Fe^\star\frac{\partial \Psi}{\partial \Fe} \Fn^{-\star}\right]
= - \frac{ \partial \Bn}{\partial \Theta}
\,.
\end{align}
\end{subequations}
Therefore, the heat equation \eqref{eq:HeatEq} may be recast as 
\begin{equation}
	\rho c_E \dot\Theta 
	= - \operatorname{Div} \mathbf Q
	+ \Theta \, \g \left[\frac{ \partial \mathbf P}{\partial \Theta} - \mathbf P \boldsymbol\alpha^{-\star} \right] \!:\! \bar{\operatorname{D}}_t\F
	+ \left[\Theta \frac{ \partial \Bn}{\partial \Theta} - \Bn \right] \!:\! \dot\Fn
	+ \rho R\,.
\end{equation}
\end{remark}

\subsubsection{Kinetic equations}

While the evolution of temperature\textemdash and consequently that of the thermal distortion $\Fth$\textemdash is known to be governed by the heat equation \eqref{eq:HeatEq}, one has yet to prescribe a constitutive model for the generalized configurational force $\Bn$ in order to complete the set of governing equations for hyper-anelasticity.
Such a constitutive model can be formulated by assuming the existence of a dissipation potential density (a Rayleigh function) $\phi=\phi(X,\Theta,\F,\Fn,\dot\Fn,\Go,\g)\,$, and the generalized configurational forces are given by
\begin{equation}\label{eq:Cnstttv_Bn}
	\Bn= -\frac{\partial \phi}{\partial \dot\Fn}\,.
\end{equation}
It is assumed that the dissipation potential is convex with respect to $\dot{\Fn}$~\citep{ziegler1958attempt, ziegler1987derivation, Germain1983, Goldstein2002ClassMech, Kumar2016}. This is equivalent to
\begin{equation}\label{eq:convex}
	\left(\frac{\partial \phi}{\partial \dot{\Fn}_2}-\frac{\partial \phi}{\partial \dot{\Fn}_1}\right)
	\!:\!\left(\dot{\Fn}_2-\dot{\Fn}_1\right) \geq 0 \,,
\end{equation}
for any $\dot{\Fn}_1$ and $\dot{\Fn}_2\,$. From \eqref{eq:Config_n} and \eqref{eq:Cnstttv_Bn}, we find a set of athermal kinetic equations in tensorial form as follows
\begin{equation}
\frac{\partial \phi}{\partial \dot\Fn} - \rho \Fe^\star\frac{\partial \Psi}{\partial \Fe} \Fn^{-\star} = 0\,.
\end{equation}
For an incompressible solid, it is modified to read
\begin{equation}
\frac{\partial \phi}{\partial \dot\Fn} - \rho \Fe^\star\frac{\partial \Psi}{\partial \Fe} \Fn^{-\star} + (\pe-\pn) \Fn^{-\star} = 0\,.
\end{equation}

\section{Concluding Remarks}
\label{Sec:Con}

In continuum mechanics, the governing equations and balance laws can be derived through various approaches. One classical method involves postulating a variational principle, where taking variations with respect to the independent fields yields the corresponding Euler-Lagrange equations. Alternatively, one can directly postulate the fundamental balance laws, such as the balance of linear and angular momenta. A third approach derives these balance laws by postulating the first law of thermodynamics (energy balance) and assuming its invariance under superimposed motions or diffeomorphisms of the ambient space. In this sense, a part of thermodynamics---namely, the first law---combined with an invariance assumption, can be used to recover the standard balance laws of continuum mechanics. The second law of thermodynamics, on the other hand, has traditionally played a different role: it serves to constrain the form of constitutive equations, ensuring their physical admissibility. That is, not all constitutive equations are thermodynamically consistent, and the second law defines the class of admissible models. In this paper, we explore how the balance laws of continuum mechanics, particularly those of hyperelasticity and anelasticity, can be derived directly from thermodynamic principles alone. We show that by postulating only the first and second laws of thermodynamics, and without assuming any invariance under superimposed motions, one can derive not only the full set of balance laws but also objectivity (material-frame-indifference).

In this paper, more specifically we present a generalization of the Coleman-Noll procedure \citep{coleman1963} to derive the balance laws of nonlinear hyperelasticity and hyper-anelasticity, including conservation of mass, balance of linear and angular momenta, and the Doyle-Ericksen formula, directly from the first and second laws of thermodynamics. Notably, this is achieved without invoking assumptions of observer invariance or pre-supposing the balance laws themselves\textemdash as it is customarily done in the classical Coleman-Noll procedure. By reframing the Clausius-Duhem inequality in terms of extended thermodynamic processes\textemdash herein defined, we demonstrate how the constitutive equations and balance laws emerge naturally within this framework.

Unlike the Green-Naghdi-Rivlin theorem \citep{Green64} (and its variants \citep{Noll1963, HuMa1977}), which suggests deriving the balance laws by postulating the invariance of the energy balance under superposed rigid body motions of the Euclidean ambient space, the procedure presented here relies solely on the first and second laws of thermodynamics. By reframing the Clausius-Duhem inequality in terms of extended thermodynamic processes, this approach eliminates the need for explicit symmetry assumptions or pre-supposed invariance principles. Notably, the Green-Naghdi-Rivlin theorem shares a conceptual foundation with Noether’s theorem, as both link symmetries to conserved quantities. In the Green-Naghdi-Rivlin framework, the invariance under rigid body motions corresponds to the translational and rotational symmetries central to Noether's theorem \citep{noether1918invariante, gunther1962einige, knowles1972class}, and the derived conservation laws—linear momentum, angular momentum, and energy—are direct outcomes of these symmetries. In contrast, the approach presented herein fundamentally differs from these ideas by deriving the constitutive equations and balance laws without reliance on symmetry or invariance.

The proposed approach not only provides a new perspective on the foundational principles of continuum mechanics but also extends the applicability of classical methods to scenarios where invariance principles are not readily defined, such as in non-Euclidean or evolving ambient spaces. This generalisation connects thermodynamics, geometry, and mechanics in a unified framework, paving the way for new insights in hyperelastic and hyper-anelastic materials.


It is worth noting that this work opens several avenues for further exploration. Notably, our thermodynamically driven derivation of balance laws bears a conceptual resemblance to Rational Extended Thermodynamics (RET), where the entropy inequality constrains constitutive structures and evolution equations without relying on imposed symmetries. 
In both frameworks, objectivity is not assumed a priori but arises from thermodynamic consistency. While our notion of an extended process differs from that used in RET---pertaining to admissible thermodynamic histories rather than an enlarged state space---the structural parallels suggest that a deeper comparison could be fruitful. Similar principles may also guide the development of other thermodynamically consistent models, particularly in systems with evolving geometries, internal variables, or microstructural effects.

\section*{Acknowledgement}

This work was partially supported by IFD -- Eurostars III Grant No. 2103-00007B and NSF -- Grant No. CMMI 1939901.

\bibliographystyle{abbrvnat}
\bibliography{ref_GenColNoll}

\section*{Appendix}

\appendix

\section{Thermal distortion}
\label{App:Thermal}

In the presence of a nonuniform temperature field $\Theta=\Theta(X,t)\,$, thermal distortion is generally incompatible \citep{stojanovic1964finite, Sadik2017Thermoelasticity}.
We denote the thermal distortion by $\Fth$; we let all the other anelastic effects be combined together in a distortion $\Fn$ such that the total anelastic distortion is written as $\Fa=\Fn\Fth\,$.
At any given point $X$ of the body, thermal expansion may be described by three stretch ratios $\{\zeta_1(X,\Theta),\zeta_2(X,\Theta),\zeta_3(X,\Theta)\}$ along three mutually independent directions forming a basis of unit vectors $\{\mathbf E_1,\mathbf E_2,\mathbf E_3\}\,$.\footnote{Note that $\{\mathbf E_K\}$ is not necessarily a coordinate basis and is not necessarily orthogonal.} The thermal distortion may be expressed as $\Fth = \sum_{K=1}^3\zeta_K \mathbf E_K \otimes \mathbf E^K$ \citep{lubarda2002multiplicative, ozakin2010geometric},
where $\{\mathbf E^K\}$ is the dual basis to $\{\mathbf E_K\}\,$.
The vector $\accentset{\scalebox{0.4}{\(\Theta\)}}{\mathbf{E}}_K=\Fth .\, \mathbf{E}_K = \zeta_K \mathbf E_K$ (with no summation over $K$) represents the thermal distortion along the unit basis vector $\mathbf{E}_K\,$. Hence, the linear thermal expansion coefficient of the material in the direction $\mathbf{E}_K$ is given by
\begin{equation}
	\alpha_K(X,\Theta)
	= \frac{1}{\|\accentset{\scalebox{0.4}{\(\Theta\)}}{\mathbf{E}}_K\|_{\mathring{\mathbf{G}}}}
	\frac{\partial}{\partial \Theta}\left[\|\accentset{\scalebox{0.4}{\(\Theta\)}}{\mathbf{E}}_K\|
	_{\mathring{\mathbf{G}}}\right]
	= \frac{1}{\zeta_K}\frac{\partial \zeta_K}{\partial \Theta}\,.
\end{equation}
It follows that $\zeta_K = \mathrm{exp}\left[\int_{\Theta_0}^\Theta \alpha_K(X,\Xi)\, d\Xi\right]\,$, where $\Theta_0=\Theta_0(X)$ is the temperature distribution such that $\zeta_K(\Theta_0)=1\,$, i.e. the temperature distribution for which the body does not experience any thermal distortion. Therefore, we may let $\zeta_K=e^{\omega_K}\,$, and the thermal distortion may be written as $\Fth = \sum_{K=1}^3 e^{\omega_K} \mathbf E_K \otimes \mathbf E^K$,
where $\omega_K=\int_{\Theta_0}^\Theta \alpha_K(X,\Xi)\, d\Xi\,$.
We construct the tensors $\boldsymbol{\alpha}$ and $\boldsymbol{\omega}$ such that their representations with respect to the basis $\{\mathbf{E}_K\}$ and its dual are given by $\boldsymbol{\alpha} = \sum_{K=1}^3 {\alpha_K} \mathbf E_K \otimes \mathbf E^K$ and $\boldsymbol{\omega} = \sum_{K=1}^3 {\omega_K} \mathbf E_K \otimes \mathbf E^K$.
One may hence write 
\begin{equation}\label{eq:partial_Fth}
	\frac{\partial\Fth}{\partial\Theta} = \sum_{K=1}^3 \frac{\partial e^{\omega_K}}{\partial\Theta} 
	\,\mathbf{E}_K \otimes \mathbf{E}^K
	= \sum_{K=1}^3 \alpha_K \,e^{\omega_K} \,\mathbf{E}_K \otimes \mathbf{E}^K
	= \boldsymbol\alpha \Fth
	= \Fth \boldsymbol\alpha
	\,,
\end{equation}
which yields
\begin{equation}
\boldsymbol\alpha = \frac{\partial\Fth}{\partial\Theta} \Fth^{-1} = \Fth^{-1} \frac{\partial\Fth}{\partial\Theta}\,.
\end{equation}

Let us denote by $[\mathrm{A}^I{}_J]$ the transformation matrix between the coordinate basis $\{\frac{\partial}{\partial X^K}\}$ and the basis $\{\mathbf E_K\}\,$, i.e. $\mathbf E_K = \mathrm{A}^I{}_K \,{\partial}/{\partial X^I}$ and $\mathbf E^K = \mathrm{A}^{-K}{}_J \,dX^J\,$, it follows that one may write
\begin{equation}
 \sum_{K=1}^3 T_K \,\mathbf E_K \otimes \mathbf E^K
	= \sum_{K=1}^3 \mathrm{A}^I{}_K \,T_K \,\mathrm{A}^{-K}{}_J \,
	\frac{\partial}{\partial X^I} \otimes dX^J\,,
\end{equation}
for any triplet of numbers $\{T_1,T_2,T_3\}\,$. In particular, the coordinate representations of $\boldsymbol\alpha$ and $\boldsymbol\omega$ in $\{X^K\}$ are hence respectively given by $\boldsymbol{\omega} = \omega^I{}_J \frac{\partial}{\partial X^I} \otimes dX^J$ and $\boldsymbol{\alpha} = \alpha^I{}_J \frac{\partial}{\partial X^I} \otimes dX^J$,
where $\omega^I{}_J = \sum_{K=1}^3 \mathrm{A}^I{}_K\, \omega_K \,\mathrm{A}^{-K}{}_J$ and $\alpha^I{}_J = \sum_{K=1}^3 \mathrm{A}^I{}_K\, \alpha_K \,\mathrm{A}^{-K}{}_J\,$. Note that it still holds in tensorial form that $\boldsymbol{\omega}(X,\Theta)=\int_{\Theta_0}^\Theta \boldsymbol{\alpha}(X,\Xi)\, d\Xi\,$.
One may write
\begin{equation}
\begin{split}
\Fth(X,\Theta) &= \sum_{K=1}^3 \zeta_K \,\mathbf E_K \otimes \mathbf E^K
	= \sum_{K=1}^3 \mathrm{A}^I{}_K \,\zeta_K \,\mathrm{A}^{-K}{}_J \,
	\frac{\partial}{\partial X^I} \otimes dX^J\\
	&= \sum_{K=1}^3 \mathrm{A}^I{}_K \,e^{\omega_K} \mathrm{A}^{-K}{}_J \,
	\frac{\partial}{\partial X^I} \otimes dX^J
	= \sum_{K=1}^3 \exp\left[\mathrm{A}^I{}_K \,\omega_K 
	\,\mathrm{A}^{-K}{}_J\right] \frac{\partial}{\partial X^I} \otimes dX^J\\
	&= \sum_{K=1}^3 \exp\left[\omega^I{}_J\right] \frac{\partial}{\partial X^I} \otimes dX^J
	= \exp\left[\boldsymbol{\omega}(X,\Theta)\right]
	\,.
\end{split}
\end{equation}
Therefore
\begin{equation}\label{eq:Fth}
\Fth(X,\Theta) 
	= \exp\left[\boldsymbol{\omega}(X,\Theta)\right]
	= \exp\left[\int_{\Theta_0}^\Theta \boldsymbol{\alpha}(X,\Xi)\, d\Xi\right]
	\,.
\end{equation}
Note that for an isothermal process, one has $\Theta(X,t)=\Theta_0(X)\,,\forall X\in \mathcal B\,,\forall t\,$, and consequently, by using \eqref{eq:Fth}, $\Fth=\mathbf I\,$, the identity tensor.

\end{document}